\documentclass[reqno,11pt]{amsart}
\usepackage{amsmath, latexsym, amsfonts, amssymb, amsthm, amscd}
\usepackage{epsfig}
\usepackage{graphics,epsf,psfrag}

\numberwithin{equation}{section}

\newtheorem{theorem}{Theorem}[section]
\newtheorem{lemma}[theorem]{Lemma}
\newtheorem{proposition}[theorem]{Proposition}

\newtheorem{rem}[theorem]{Remark}

\DeclareMathOperator{\sign}{\mathrm{sign}}

\newcommand{\ind}{\mathbf{1}}

\newcommand{\R}{\mathbb{R}}
\newcommand{\Z}{\mathbb{Z}}
\newcommand{\N}{\mathbb{N}}
\renewcommand{\tilde}{\widetilde}

\newcommand{\cA}{{\ensuremath{\mathcal A}} }

\newcommand{\cF}{{\ensuremath{\mathcal F}} }

\newcommand{\cN}{{\ensuremath{\mathcal N}} }
\newcommand{\cL}{{\ensuremath{\mathcal L}} }

\newcommand{\cD}{{\ensuremath{\mathcal D}} }
\newcommand{\cZ}{{\ensuremath{\mathcal Z}} }

\newcommand{\bP}{{\ensuremath{\mathbf P}} }
\newcommand{\bE}{{\ensuremath{\mathbf E}} }

\DeclareMathSymbol{\leqslant}{\mathalpha}{AMSa}{"36} 
\DeclareMathSymbol{\geqslant}{\mathalpha}{AMSa}{"3E} 
\DeclareMathSymbol{\eset}{\mathalpha}{AMSb}{"3F}     
\newcommand{\dd}{\,\text{\rm d}}             
\DeclareMathOperator*{\inter}{\bigcap}       
\newcommand{\sumtwo}[2]{\sum_{\substack{#1 \\ #2}}} 

\newcommand{\bbE}{{\ensuremath{\mathbb E}} }

\newcommand{\bbL}{{\ensuremath{\mathbb L}} }

\newcommand{\bbP}{{\ensuremath{\mathbb P}} }

\newcommand{\ga}{\alpha}
\newcommand{\gb}{\beta}

\newcommand{\gep}{\varepsilon}       
\newcommand{\gp}{\varphi}

\newcommand{\gD}{\Delta}

\newcommand{\go}{\omega}
\newcommand{\gO}{\Omega}
\newcommand{\gl}{\lambda}

\newcommand{\gs}{\sigma}
\newcommand{\gS}{\Sigma}

\makeatletter
\def\captionfont@{\footnotesize}
\def\captionheadfont@{\scshape}

\long\def\@makecaption#1#2{%
  \vspace{2mm}
  \setbox\@tempboxa\vbox{\color@setgroup
    \advance\hsize-6pc\noindent
    \captionfont@\captionheadfont@#1\@xp\@ifnotempty\@xp
        {\@cdr#2\@nil}{.\captionfont@\upshape\enspace#2}%
    \unskip\kern-6pc\par
    \global\setbox\@ne\lastbox\color@endgroup}%
  \ifhbox\@ne 
    \setbox\@ne\hbox{\unhbox\@ne\unskip\unskip\unpenalty\unkern}%
  \fi
  \ifdim\wd\@tempboxa=\z@ 
    \setbox\@ne\hbox to\columnwidth{\hss\kern-6pc\box\@ne\hss}%
  \else 
    \setbox\@ne\vbox{\unvbox\@tempboxa\parskip\z@skip
        \noindent\unhbox\@ne\advance\hsize-6pc\par}%
\fi
  \ifnum\@tempcnta<64 
    \addvspace\abovecaptionskip
    \moveright 3pc\box\@ne
  \else 
    \moveright 3pc\box\@ne
    \nobreak
    \vskip\belowcaptionskip
  \fi
\relax
}
\makeatother
\def\writefig#1 #2 #3 {\rlap{\kern #1 truecm
\raise #2 truecm \hbox{#3}}}

\newcommand{\tf}{\textsc{f}}

\newcommand{\M}{\textsf{M}}

\begin{document}

\title[Copolymers at selective interfaces]{
A numerical approach to
\\
 copolymers at selective interfaces
}

\author{Francesco Caravenna}

\address{Universit\`a di Milano-Bicocca, Dipartimento di Matematica e Applicazioni, \mbox{Edificio} U5,\break via Cozzi 53, 20125 Milano, Italy \hfill \break
\indent \textit{and} \hfill\break
\indent Laboratoire de Probabilit{\'e}s de P 6 \& 7 (CNRS U.M.R. 7599)
and  Universit{\'e} Paris 7 -- Denis Diderot,
U.F.R. Mathematiques, Case 7012, 2 place Jussieu, 75251 Paris cedex 05, France}

\email{f.caravenna\@@sns.it}

\author{Giambattista Giacomin}
\address{Laboratoire de Probabilit{\'e}s de P 6\ \& 7 (CNRS U.M.R. 7599)
  and  Universit{\'e} Paris 7 -- Denis Diderot,
U.F.R.                Mathematiques, Case 7012,
                2 place Jussieu, 75251 Paris cedex 05, France
\hfill\break
\indent{\it Home page:}
{\tt http://www.proba.jussieu.fr/pageperso/giacomin/GBpage.html}}
\email{giacomin\@@math.jussieu.fr}

\author{Massimiliano Gubinelli}
\address{Dipartimento di Matematica Applicata ``U.~Dini'',
Universit\`a di Pisa, via Bonanno Pisano 25b, 56126 Pisa, Italy}
\email{m.gubinelli\@@dma.unipi.it}

\date{\today}

\begin{abstract}
We consider a model of a {\sl random copolymer at a selective interface} which undergoes a localization/delocalization transition. In spite of the several rigorous results  available for this model, the theoretical characterization of the phase transition has remained elusive and there is still no agreement about several important issues, for example the behavior of the polymer
near the phase transition line. From a rigorous viewpoint non coinciding upper and lower bounds on the critical line are known.

In this paper we combine numerical computations with rigorous arguments to get to a better understanding of the phase diagram. Our main results include:
\\
-- Various numerical observations  that suggest  that the critical line lies
\emph{strictly} in between the two  bounds.
\\
-- A rigorous statistical test based on concentration inequalities and super--additivity,
for determining  whether a given point of the phase diagram is in the localized phase. This is applied
in particular to show that, with a very low level of error, the lower bound does not coincide with the critical line.
\\
-- An analysis of the precise asymptotic behavior of the partition function in the delocalized phase, with particular attention to the effect of rare {\sl atypical} stretches
in the disorder sequence   and on whether or not in the delocalized regime
the polymer path has a Brownian scaling.
\\
-- A new proof of the lower bound on the critical line. This proof
relies on a characterization of the localized regime which is more appealing for
interpreting the numerical data.
\end{abstract}

\maketitle

\section{Introduction}
\label{sec:intro}
\setcounter{equation}{0}

\subsection{The model}

\begin{figure}[h]
\begin{center}
\leavevmode
\epsfysize =5 cm
\psfragscanon
\psfrag{0}[c][l]{$0$}
\psfrag{n}[c][l]{ $n$}
\psfrag{S}[c][l]{$S_n$}
\epsfbox{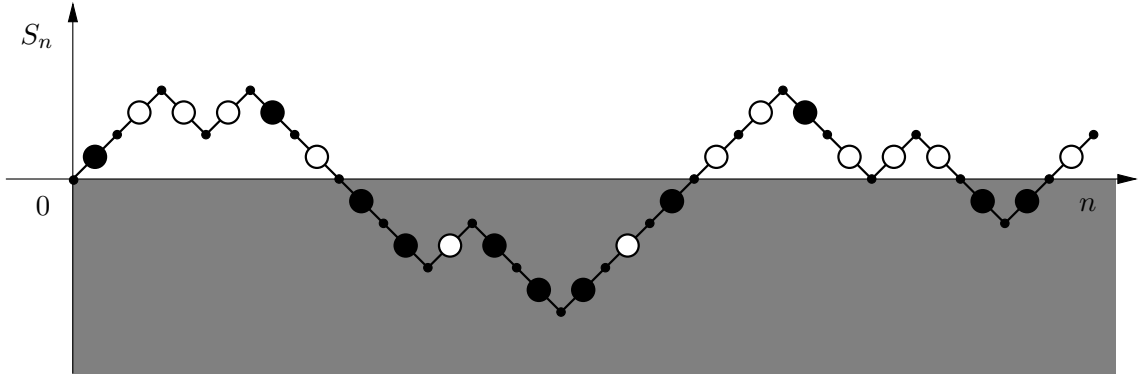}
\end{center}
\caption{\label{fig:cop_bonds}
The process we have introduced
is a model for a non--homogeneous polymer, or {\sl copolymer},
near an interface, the horizontal axis, between two selective solvents, say oil (white) and water (grey).
 In the drawing the monomer {\sl junctions} are the small black rounds and the monomers are the bonds of the random walk. The big round in the middle of each monomer
gives the sign of the charge  (white = positive charge = hydrophobic monomer, black = negative charge = hydrophilic monomer). When $h > 0$ water is  the unfavorable solvent and the question
is whether the polymer is {\sl delocalized} in oil or if it is still more profitable
to place a large number of monomers in the preferred solvent, leading in such a way
to the {\sl localization at the interface} phenomenon.
The conventional  choice of $\sign(0)$ we have made reflects the fact that
the charge is assigned to bonds rather than points.  }
\end{figure}

Let $S=\{S_n \}_{n=0,1,\ldots}$ be a random walk
with $S_0=0$ and $S_n=\sum_{j=1}^n X_j$, $\{ X_j\}_j$ a sequence
of IID random variables and $\bP \left( X_1=  1\right)=\bP \left( X_1=  -1\right)=1/2$.
For $\gl\ge 0$,  $h \ge 0$, $N\in 2\N$  and
$\go =\{ \go_j\}_{j=1,2, \ldots} \in \R ^\N$
we introduce the probability measure $\bP _{N, \go}^{\gl,h}$ defined by
\begin{equation}
\label{eq:Boltzmann}
\frac {\dd \bP_{N, \go}^{\gl, h}} {\dd \bP} (S)
\, = \,
\frac 1
{{\tilde Z}_{N,\go}^{\gl,h}}
{\exp\left( \gl \sum_{n=1}^N \left( \go_n +h\right) \sign \left(S_n\right)
\right)}
,
\end{equation}
where ${\tilde Z}_{N,\go}^{\gl, h}$ is the partition function and
$\sign \left(S_{2n}\right)$ is set to be equal to $\sign \left(S_{2n-1}\right)$
for any $n$ such that $S_{2n}=0$. This is a natural choice, as it is explained in the caption of
Fig.~\ref{fig:cop_bonds}.

\smallskip

For what concerns the {\sl charges} $\go$  we put ourselves in a
 {\sl quenched set--up}:
$\go$ is a typical realization
of an IID sequence of
 random variables (we denote by $\bbP$ its law).
We suppose that
\begin{equation}
\label{eq:hypM}
\M (\ga) := \bbE\left[ \exp\left( \ga \go_1 \right)\right]<\infty \, ,
\end{equation}
for every $\ga$ and that $\bbE\left[ \go_1 \right]=0$.
Moreover we fix $\bbE[ {\go_1}^2]=1$.

\subsection{The free energy and the phase diagram}

We introduce the free energy of the system
\begin{equation}
\label{eq:free_energy}
f(\gl ,h)\, = \, \lim_{N\to \infty} \frac 1N \log {\tilde Z}^{\gl, h}_{N,\go}.
\end{equation}
The limit has to be understood in the $\bbP\left( \dd \go \right)$--almost
sure sense, or
 in the $\bbL_1\left( \bbP\right)$
sense, and $f (\gl,h)$ does not depend on $\go$.
 A  proof of the existence of such a limit goes along a standard superadditive argument
and we refer to \cite{cf:G} for the details, see however \S~\ref{sec:superadd} below.
By convexity arguments one easily sees that the free energy is a continuous function.

We observe that
\begin{equation}
\label{eq:delocfe}
f(\gl , h) \, \ge \, \gl h.
\end{equation}
In fact if we set $\gO_N^+= \{S:\, S_n>0$ for $ n=1, 2, \ldots , N\}$
\begin{multline}
\label{eq:step_deloc}
\frac 1N \log {\tilde Z}_{N,\go}^{\gl, h}  \ge
\frac 1N \log \bE
\left[
\exp\left(
\gl \sum_{n=1}^N
\left( \go_n +h\right) \sign \left(S_n\right)
\right)
;  \gO_N^+
\right]
\\
= \frac {\gl} {N} \sum_{n=1}^N \left( \go_n +h\right) \, + \,
\frac 1N
\log \bP \left( \gO_N^+\right)\, \stackrel{N \to \infty}{\longrightarrow}\, \gl h ,
\end{multline}
where the limit has to be understood in the $\bbP(\dd \go)$--almost sure sense:
notice that
we have  used the law of large numbers.
We have of course also applied the
well known fact that $\bP\left( \gO_N^+\right)$ behaves like
$N^{-1/2}$ for $N$ large \cite[Ch. III]{cf:Feller}. In view of \eqref{eq:free_energy} and of \eqref{eq:step_deloc}
we partition the phase diagram in the following way:

\smallskip
\begin{itemize}
\item The localized region: $\cL = \left\{ (\gl , h): \, f(\gl, h)-\gl h>0\right\}$;
\item The delocalized region:  $\cD = \left\{ (\gl , h): \, f(\gl, h)- \gl h=0\right\}$.
\end{itemize}
\smallskip

\smallskip

This phase diagram decomposition does correspond to different
behaviors of the trajectories of the copolymer: we will come back to this
important issue in  \S~\ref{sec:pathwise}.

\smallskip

We sum up in the following theorem what is known about
the phase diagram of the model.

\medskip

\begin{theorem}
\label{th:sumup}
There
exists a continuous increasing function $h_c: \, [0, \infty) \longrightarrow
[0, \infty)$, $h_c(0)=0$, such that
\begin{equation}
\cL \, = \, \left\{ (\gl, h) :\, h < h_c (\gl) \right\}
\ \ \text{ and } \ \
\cD \, = \, \left\{ (\gl, h) :\, h \ge h_c (\gl) \right\}.
\end{equation}
Moreover
\begin{equation}
\label{eq:sumupq}
\underline h (\gl) \, :=\,
\frac 1{4\gl /3} \log \emph\M \left( -4\gl /3\right)\le
h_c (\gl)\le \frac 1{2\gl } \log \emph\M \left( -2\gl \right)
\, =: \, \overline{h} (\gl).
\end{equation}
This implies that the slope at the origin belongs to $[{2}/{3},1]$,
in the sense  that the inferior limit of $h_c(\gl)/\gl$ as $\gl \searrow 0$
is not smaller than $2/3$ and the superior limit is not larger than $1$.
\end{theorem}

\medskip

\begin{rem}\rm
In~\cite{cf:GT} it is proven that the limit of $h_c(\gl)/\gl$ as $\gl \searrow  0$ does exist and
it is independent of the distribution of $\go_1$, at least
when $\go_1$ is a bounded symmetric random variable or when $\go_1 $
is a standard Gaussian variable \cite{cf:GT}. 
This {\sl universal character} of the slope at the origin makes this quantity
very interesting.
\end{rem}

\medskip

Theorem~\ref{th:sumup} is a mild generalization of the
results proven
in \cite{cf:BdH} and \cite{cf:BG2}:
the extension lies in the fact that $\go_1$ is not necessarily symmetric
and a proof of it requires minimal changes with respect to the
arguments in  \cite{cf:BG2}.
The  lower bound in \eqref{eq:sumupq} is actually
proven explicitly in Appendix~\ref{app:prooflb} (see also Section~\ref{sec:lb}),
but we stress that we present
this proof because it is a new one and because it gives
some insight on the computational results.
For what follows we set
\begin{equation}
\label{eq:hm}
h^{(m)}(\gl) \, = \, \frac 1{2m\gl } \log \M\left( -2m\gl \right),
\end{equation}
for $m >0$. Observe that the curves $\underline h(\cdot)$ and $\overline h(\cdot)$ defined in \eqref{eq:sumupq} correspond respectively to $m=2/3$ and $m=1$, and that $\frac{\dd}{\dd\gl} h^{(m)}(\gl) |_{\gl=0} = m$.

\medskip

\begin{rem}
\label{rem:Z}
\rm
Notice that one can write
\begin{equation}
\label{eq:Boltzmann1}
\frac {\dd \bP_{N, \go}^{\gl, h}} {\dd \bP} (S)
\, = \,
\frac 1
{{Z}_{N,\go}^{\gl,h}}
{ \exp\left( -2 \gl \sum_{n=1}^N \left( \go_n +h\right) \Delta_n
\right)}
,
\end{equation}
with $\Delta_n = \left(1-\sign (S_n)\right)/2$ and $Z_{N,\go} := Z_{N,\go}^{\gl, h}$
a new partition function which coincides with $\tilde Z_{N, \go} \exp \left( - \gl \sum_{n=1}^N (\go_n +h) \right)$
and therefore
we have
\begin{equation}
\label{eq:felim2}
\tf (\gl, h) \, := \, \lim_{N \to \infty}
\frac 1N \log Z_{N, \go}= f(\gl, h)-\gl h.
\end{equation}
This limit of course has to be interpreted in the $\bbP (\dd \go)$--a.s. sense.
We stress that,
even if  equivalent to $\tilde Z_{N, \go}\exp(-\gl h N)$ in the Laplace asymptotic sense,
$Z_{N , \go}$ turns out to be substantially more useful. This had been
already realized in \cite{cf:BdH}, but
for our results    looking at $Z_{N, \go}$, rather
than  $\tilde Z_{N, \go}$, is even more essential. Moreover from now on
$\tf (\gl, h)$, rather than $f(\gl , h)$,  will be for us the free energy.
 \end{rem}

We will use repeatedly also the partition function associated to the
model {\sl pinned} at the right endpoint:
\begin{equation}
\label{eq:pinned}
Z_{N, \go}^{\gl, h}(x) \, :=\,
\bE\left[
 \exp\left(
 -2 \gl \sum_{n=1}^N \left( \go_n +h\right) \Delta_n
\right); \, S_N= x
\right].
\end{equation}
It is worth recalling that one can substitute $Z_{N, \go}^{\gl, h}$ with $Z_{N, \go}^{\gl, h}(x)$,
any fixed even $x$ (recall that $N \in 2\N$), in
\eqref{eq:felim2} and the limit is unchanged, see e.g. \cite{cf:BdH} or \cite{cf:G}.

 \subsection{A random walk excursions viewpoint}
 \label{sec:notation}
 We present here a different viewpoint on the process:
 this turns out to be useful for the intuition and it will be used
 in some technical steps.

 We call  $\eta$  the first
return time of the walk $S$ to $0$, that is $\eta:=\inf\left\{  n\ge
1:S_{n}=0\right\} $, and  set $K(2n):=\bP \left(\eta=2n\right)$ for $n\in\N$.
It is well known that $K(\cdot)$ is decreasing on the even natural numbers and
\begin{equation}
\label{eq:asympt}
\lim_{x\in 2\N, x\to \infty} x^{3/2} K(x)=
\sqrt{2/\pi },
\end{equation}
see e.g. \cite[Ch. 3]{cf:Feller}.
Let the IID sequence $\left\{ \eta_j \right\}_{j=1,2, \ldots}$
denote the inter--arrival times at $0$ for $S$, and we set $\tau_k := \eta_0 + \ldots + \eta_k$.
 If we introduce also
 $\ell _N =\max\{j \in \N \cup \{0\} :
\tau_j \le N\}$, then by  exploiting
the up--down symmetry of the excursions of $S$  we directly obtain
\begin{equation}
\begin{split}
\label{eq:reducttoexc}
& Z_{N,\go} (0) \, =\, \bE \left[ \prod_{j=1}^{\ell_N} \varphi \Bigg(\gl \sum_{n=\tau_{j-1}+1}^{\tau_j} \go_n + \gl h  \eta_j \Bigg) ; \tau_{\ell_N} = N \right] \\
    & \quad =\, \sum_{l=0}^N \sumtwo{x_0, \ldots, x_l \in 2\N}{0=:x_0 < \ldots < x_l:= N} \prod_{i=1}^l \; \varphi \Bigg(\gl \sum_{n=x_{i-1}+1}^{x_i} \go_n + \gl h  (x_i - x_{i-1}) \Bigg) \; K(x_i - x_{i-1}) \,,
\end{split}
\end{equation}
with $\varphi (t) : =   \left( 1+\exp(-2t)\right) / 2$.
Of course the formula for $Z_{N,\go}$
is just  slightly different.

Formula \eqref{eq:reducttoexc} reflects the fact that what really
matters for the copolymer are the return times to the interface.

\subsection{Known and conjectured  path properties}
\label{sec:pathwise}
The question of whether splitting the phase diagram
into the regions $\cL$ and $\cD$ does correspond
to really different path behaviors has a positive answer,
at least if we do not consider the critical case, that is
if we consider the path behavior for $(\gl, h) \in\cL$ and
for $(\gl, h)$ in the interior of $\cD$.
However, while the localized regime is rather well understood,
the delocalized one remains somewhat elusive (we take up this point again in Section~\ref{sec:path1}).
More precisely:
\smallskip
\begin{itemize}
\item For $(\gl, h) \in\cL$ one knows that the polymer is going to
stay very close to the interface, {\sl essentially} at distance $O(1)$ and the polymer
becomes positive recurrent for $N\to \infty$.
Due to the disordered distribution of the charges, even the most
elementary  results in this direction require a careful formulation and
we prefer to refer to \cite{cf:AZ}, \cite{cf:BisdH} and \cite{cf:Sinai}.
\item
For $(\gl, h)$ in the interior of $\cD$
one can prove by large deviation arguments that
there are $o(N)$ visits to the unfavorable solvent
and by more sophisticate arguments that these visits are actually $O(\log N)$
 \cite{cf:GT}. These results are
  in sharp contrast with what happens in $\cL$ and in this sense
they are satisfactory. However they give  at the same time still a  weak information on the paths,
above all if compared to what is available for non disordered models, see e.g.
\cite{cf:MGO}, \cite{cf:DGZ}, \cite{cf:CGZ} and references therein, namely Brownian scaling,
which in turn is a consequence of the fact that all the visits in the unfavorable
solvent happen very close to the boundary points, that is the origin, under the measure
$\bP _{N, \go}^{\gl,h}$. In non disordered models one can in fact prove that
the polymer becomes transient and that it visits the unfavorable solvent, or any point below a fixed
level,  only a finite
number of times. Recently it has been shown  \cite{cf:GT}
that such a result cannot hold as stated, at least for $h <\overline{h}(\gl)$, for the
disordered copolymer. However the results in  \cite{cf:GT}
leave open the possibility of Brownian scaling in the whole delocalized region.
\end{itemize}
 \smallskip

\subsection{Outline of the results}
\label{sec:Outline}

Formula \eqref{eq:sumupq} leaves an important gap, that
hides the only partial understanding of the nature of this delocalization/localization
transition.
Our purpose is to go toward filling this gap: our results are both of theoretical and numerical nature.
At the same time we address the delocalization issues raised
in \S~\ref{sec:pathwise}, which are intimately related with the precise
asymptotic behavior of $Z_{N,\go}$ and of  $Z_{N,\go}(0)$.
More precisely:

 \medskip
 \begin{enumerate}
 \item In Section~\ref{sec:testing} we present a statistical test with explicit error bounds, see Proposition~\ref{th:stat}, based on
 super--additivity and concentration inequalities, to state that a point $(\gl, h)$ is localized. We apply this test to show that, with a very low level of error, the lower bound $h=\underline h(\gl)$ defined in \eqref{eq:sumupq} does not coincide with the critical line.
 \item In Section~\ref{sec:lb} we give the outline of a new proof of the main
 result in \cite{cf:BG2}. The details of the proof are in Appendix B and we point out
 in particular Proposition~\ref{prop:stopping1}, that gives a necessary and sufficient
 condition for localization. This viewpoint on the transition, derived from
 \cite[Section~4]{cf:GT}, helps substantially in interpreting the {\sl irregularities} in the
 behavior  of $\left\{ Z_{N, \go}\right\}_N$ as $N \nearrow \infty$.
 \item In Section~\ref{sec:path} we pick up the conjecture of Brownian scaling
 in the delocalized regime both in the intent of testing it and
 in trying to asses with reasonable confidence that $(\gl, h)$ is in the interior
 of $\cD$. In particular, we present quantitative evidences in favor of the fact that the upper bound $h= \overline h (\gl)$ defined in \eqref{eq:sumupq} is strictly greater than the critical line. We stress that this is a very delicate issue, since delocalization, unlike
 localization, does not appear to be reducible to a finite volume issue.
 \item Finally, in Section~\ref{sec:guess}, we report the results of a numerical attempt
 to determine the critical curve. While this issue has to be treated with care,
 mostly for the reasons raised in point 4 above, we observe
 a surprising phenomenon: the critical curve appears to be  very close
 to $h^{(m)}(\cdot)$ for a suitable value of $m$. By the universality
 result proven in \cite{cf:GT}, building on the free energy Brownian scaling result proven
 in \cite{cf:BdH}, the slope at the origin of $h_c(\cdot)$ does not depend on the law of $\go$.
 Therefore if really $h^{(m)}(\cdot)= h_c (\cdot)$, since the slope at the origin
 of $h^{(m)}(\cdot)$ is $m$,
 $m$ is the  universal constant we are looking for.
 We do not believe that the numerical evidence
 allows to make a clear cut  statement, but what we observe is compatible
 with such a possibility.
 \end{enumerate}

We point out that our numerical results are based on a numerical computation of the partition function $Z_{N,\go}$, exploiting the standard transfer--matrix approach (this item is discussed in more details in Appendix~\ref{app:algo}).

\subsection{A quick  overview of the literature}
\label{sec:overv}

The {\sl copolymer in the proximity of an interface} problem
has a long literature, but possibly the first article that attracted the attention
of mathematicians is \cite{cf:GHLO}.
Here we are going to
focus on very specific issues and the most interesting for our purposes
is that in the physical literature both the conjecture that $\underline{h}(\cdot)=
h_c (\cdot)$ (cf. \cite{cf:Monthus} and \cite{cf:SSE})  and that $\overline{h}(\cdot)=
h_c (\cdot)$  (cf. \cite{cf:TM}) are set forth.
The approaches are non rigorous, mostly based on replica computations,
with the exception of \cite{cf:Monthus} whose method is the real space renormalization
technique for one--dimensional disordered systems first proposed in \cite{cf:Fisher}
in the context of quantum Ising model with transverse field
and then applied with remarkably precise results to random walk in random environment,
see e. g. \cite{cf:LDMF99}.
The result in \cite{cf:BG2}, that $\underline{h} (\cdot) \le h_c (\cdot)$,
is obtained by exploiting the path behavior of the copolymer near criticality
suggested in   \cite{cf:Monthus}. This strategy may by summed up by:
the localized polymer close to criticality is mostly delocalized
in the upper half--plane and it keeps in the lower half--plane
only the rare portions with an atypically negative charge.
The numerical results that we set forth in this work are saying that
this strategy is not good enough.

\smallskip

At the opposite end, the result $\overline{h}(\cdot)\ge h_c(\cdot)$, albeit
relatively subtle,
is absolutely elementary to prove \cite{cf:BdH}. And such a bound does not depend at all
on the details of the walk: any non trivial null
recurrent walk with increments in $\{-1, 0, +1\}$ leads to the
same upper bound. This suggests that such a bound
is too rough. One can however prove that the standard procedure
for obtaining upper bounds that goes under the name of
{\sl constrained annealing} cannot improve such a bound \cite{cf:CG}.
This is in any case far from being a proof that $\overline{h}(\cdot)= h_c(\cdot)$,
and in fact the numerics suggest that this is not the case.

\smallskip

In the literature one finds also a large number of numerical works on copolymers,
we mention here for example \cite{cf:CW}, \cite{cf:SW} and references therein.
As far as we have seen, the attention is often shifted toward different aspects,
notably of course the  issue of critical exponents, and the more
complex model in which the polymer is not directed but rather self--avoiding,
see   \cite{cf:CW} and \cite{cf:SW} also for some rigorous results and references in such a direction.

Our work has been led rather by the idea that understanding
the precise location of the critical curve is a measure of our understanding of the nature of the transition. Understanding that, in turn, could promote an advance on the mathematical analysis of the copolymer and, more generally,  of this kind of disordered models.


\section{A statistical test for the localized phase}
\label{sec:testing}

\subsection{Checking localization at finite volume}
\label{sec:superadd}
At an intuitive level one is led to believe that, when the copolymer is localized,
it should be possible to detect it
by looking at the system before the infinite volume limit.
This intuition is due to the fact that in the localized phase
the length of each excursion is finite, therefore for $N$ {\sl much larger }
that the {\sl typical} excursion length one should already observe the
localization phenomenon in a quantitative way.
The system being disordered of course does not help,
because it is more delicate to make sense of what
{\sl typicality} means in a non translation invariant set--up.
 However the translation invariance can be recovered
 by averaging and in fact it turns out to be rather easy to give
 a precise meaning to the intuitive idea we have just mentioned.
 The key word here is super--additivity of the averaged free energy.
 \smallskip

 In fact by considering only the $S$ trajectories such that $S_{2N}=0$ and by applying
 the Markov property of $S$ one directly verifies
  that for any $N, M\in \N$
 \begin{equation}
 \label{eq:superadd}
 Z_{2N+2M, \go} (0)\, \ge \, Z_{2N, \go}(0)\, Z_{2M, \theta^{2N} \go}(0),
 \end{equation}
 $(\theta \go)_n=\go_{n+1}$,
 and therefore
 \begin{equation}
 \left\{ \, \bbE\log Z_{2N, \go} (0)
 \right\}_{N=1,2, \ldots}
  \end{equation}
  is a super--additive sequence, which immediately entails
  the existence of the limit of $ \bbE[\log Z_{2N, \go}(0)]/2N$ and the fact that
  this limit coincides with the supremum of the sequence.
  Therefore from the existence of the quenched free energy we have that
  \begin{equation}
  \tf(\gl, h) \, =\, \sup_N \frac 1{2N} \bbE\log Z_{2N, \go}(0)\,.
  \end{equation}
In a more suggestive way one may say that:
\begin{equation}
\label{eq:loc_char}
(\gl, h) \in \cL \  \  \Longleftrightarrow  \ \
\text{there exists } N\in \N \ \, \text{such that} \,  \  \bbE\log Z_{2N, \go}(0)>0\,.
\end{equation}
The price one pays for working with a disordered system
is precisely in taking the $\bbP$--expectation
and from the numerical viewpoint it is an heavy price:
even with the most positive attitude one cannot expect
to have access to  $\bbE\log Z_{2N, \go}(0)$
by direct numerical computation for $N$ above $10$.
Of course in principle small values of $N$ may suffice
(and they do in some cases, see Remark~\ref{rem:N=2}), but they do not
suffice to tackle the specific issue we are interested in.
We elaborate at length on this interesting issue in \S~\ref{sec:computer_assisted}.
\smallskip

\begin{rem}
\label{rem:N=2}
\rm
An elementary application of the localization criterion \eqref{eq:loc_char} is
obtained for $N=1$: $(\gl, h)\in \cL$ if
\begin{equation}
\label{eq:N=2}
\bbE\left[
\log\left(
\frac 12+ \frac 12 \exp\left(
-2\gl \left(\go_1+ \go_2+2h\right)
\right)
\right)
\right]>0.
\end{equation}
In the case $\bbP (\go_1=\pm 1)=1/2$ from
\eqref{eq:N=2}
we obtain that for $\gl$ sufficiently large
$h_c(\gl) > 1- c/\gl$, with $c= (1/4)\log(2\exp(4) -1)\approx1.17$.
From $\underline{h}(\cdot)$ we obtain the same type of bound,
with $c=(3/4)\log 2\approx 0.52$. This may raise some hope that
for $\gl$ large an explicit, possibly computer assisted, computation
for small values of $N$ of $\bbE \log Z_{2N, \go}(0)$ could
lead to new estimates. This is not the case, as we show in~\S~\ref{sec:computer_assisted}.
\end{rem}

\subsection{Testing by using concentration}
In order to decide whether  $\bbE\log Z_{2N, \go}(0)>0$ we
resort to a Montecarlo evaluation of $\bbE\log Z_{2N, \go}(0)$
that can be cast into a statistical test with explicit error bound
by means of concentration of measure ideas.
This procedure is absolutely general, but we have to choose
a set--up for the computations and we take the simplest:
 $\bbP(\go_1=+1)=\bbP(\go_1=-1)=1/2$.
 The reason for this choice is twofold:
 \begin{itemize}
 \item if $\go_1$ is a bounded random variable, a Gaussian concentration
 inequality holds and if $\go$ is symmetric and it takes only two values
 then one can improve on the explicit constant in such an inequality.
 This speeds up in a non negligible   way the computations;
 \item generating {\sl true randomness}
is out of reach, but playing head and tail is certainly
the most elementary case in such a far reaching task (the random numbers
issue is briefly discussed in Appendix~\ref{app:algo} too).
 \end{itemize}

\smallskip
A third reason to restrict testing to the Bernoulli case is explained at the end of the
caption of Table \ref{tbl:2}.
\smallskip

We start the testing procedure by stating the null hypothesis:
\begin{equation}
\label{eq:H0}
\text{H}0: \  \ \bbE\log Z_{2N, \go}(0)\le 0.
\end{equation}
$N$ in H0  can be chosen
arbitrarily. We stress that refusing H0 implies $\bbE\log Z_{2N, \go}(0)>0$,
which by \eqref{eq:loc_char} implies localization.

The following concentration inequality for Lipschitz functions holds
for the uniform measure on $\{-1,+1\}^N$: for every
function $G_N:\{-1,+1\}^N \to \R$
such that $\vert G_N(\go)-G_N(\go^\prime )\vert
\le C_{\text{Lip}} \sqrt(\sum_{n=1}^N (\go_n-\go^\prime_n)^2)$, where
$C_{\text{Lip}}$ a positive constant
and $G_N(\go)$ is an abuse of notation for $G_N(\go_1, \ldots, \go_N)$, one has
\begin{equation}
\label{eq:concentration}
\bbE\left[\exp\left(\ga \left(G_N(\go)- \bbE[G_N(\go)]\right)\right)\right]\, \le\,
\exp\left(\ga^2C_{\text{Lip}}^2\right),
\end{equation}
for every $\ga$.
Inequality \eqref{eq:concentration} with an extra factor $4$ at the exponent
can be extracted from the proof of Theorem 5.9, page 100 in \cite{cf:Ledoux}.
Such an inequality holds for variables taking values in $[-1,1]$:
the factor $4$ can be removed for the particular case we are considering
(see \cite[p. 110--111]{cf:Ledoux}).
In our case $G_N(\go)= \log Z_{2N,\go}(0)$. By applying the
 Cauchy--Schwarz inequality one obtains that $G_N$ is Lipschitz with
$C_{\text{Lip}}= 2\gl\sqrt{N}$.
Let us now consider an IID sequence  $\{ G^{(i)}_N(\go)\}_i$
with $G^{(1)}_N (\go)=G_N(\go)$: if H0 holds then we have that for every $n\in \N$, $u>0$
and $\ga= un/8\gl^2N$
\begin{equation}
\begin{split}
\bbP\left(
\frac 1n \sum_{i=1}^n G^{(i)}_N(\go) \ge u
\right)
\, &\le  \,
\bbE\left[
\exp\left(
\frac\ga n \left(G_N(\go)- \bbE[G_N(\go)]\right)
\right)
\right]^n
\exp\left(-\ga\left( u- \bbE[G_N(\go)]\right)\right)
\\
&\le \,
\exp\left(\frac{4 \ga^2 \gl^2N}{n}-\ga u\right)
\\
&= \, \exp\left( -\frac{u^2 n}{16 \gl^2 N}\right) .
\end{split}
\end{equation}

Let us sum up what we have obtained:

\medskip
\begin{proposition}
\label{th:stat}
Let us call $\widehat u_n$ the average of a sample of $n$ independent
realizations of $\log Z_{2N,\go}^{\gl,h}(0)$. If $\widehat u_n>0$ then we may
refuse H0, and therefore  $(\gl,h)\in \cL$, with a level
of error not larger than $\exp\left( -{\widehat u_n ^2 n}/{16 \gl^2 N}\right)$.
\end{proposition}
\medskip

\subsection{Numerical tests}
We
 report in Table \ref{tbl:1}
 the most  straightforward application
of Proposition \ref{th:stat}, obtained by a numerical computation of $\log Z_N$ for a sample of~$n$ independent environments~$\go$.
We aim
at seeing how far above $\underline{h}(\cdot)$ one can go
and still claim localization,
 keeping a reasonably small probability of error.
\begin{table}[h]
\begin{center}
\begin{tabular}{|c|c|c|c|}
\hline
$\gl $ &$0.3$&$0.6$&$1$\\
\hline
$h$&  0.22 &0.41&0.58\\
\hline
$p$--value& $1.5\times 10^{-6}$&$9.5\times 10^{-3}$&$1.6 \times 10^{-5}$\\
\hline
$\underline{h}(\gl)$& 0.195 &0.363&0.530\\
\hline
$\overline{h}(\gl)$
 & 0.286 &0.495&0.662\\
\hline
$N$& 300000 &500000&160000\\
\hline
$n$& 225000 &330000&970000\\
\hline
C. I.  99\% & $7.179\pm0.050$ &$9.011\pm 0.045$& $7.643 \pm 0.025$\\
\hline
\end{tabular}
\end{center}
\caption{\label{tbl:1}
According to our numerical computations,
the three pairs $(\gl,h)$ are in $\cL$ and this has been tested with the stated
 $p$--values (or probability/level of error).
We report the values of $\overline{h}(\gl)$ and $\underline{h}(\gl)$ for reference.
Of course in these tests there is quite a bit of freedom in the choice of $n$
and $N$: notice that $N$ enters in the evaluation of the $p$--value also
because a larger value of $N$ yields a larger value of $\bbE \log Z_{2N,\go}^{\gl,h}(0)$.
In the last line we report  standard Gaussian $99\%$ confidence intervals for
$\bbE \log Z_{2N,\go}^{\gl,h}(0)$. Of course the $p$--value under the Gaussian assumption
turns out to be totally negligible. }
\end{table}

\begin{rem}
\rm
One might be tempted to interpolate between the values in Table  \ref{tbl:1},
or possibly to get results for small  values of $\gl$ in order to
extend the result of the test to the slope of the critical curve in the origin.
However the fact that $h_c(\gl)$ is strictly increasing does not help much in this direction
and the same is true for
 the finer result, proven in \cite{cf:BG}, that $h_c(\gl)$ can be written as $U(\gl)/\gl$,
$U(\cdot)$ a convex function.
\end{rem}

\subsection{Improving on $\underline{h}(\cdot)$ is uniformly hard}
\label{sec:computer_assisted}
One can get much smaller $p$--values at little computational cost  by choosing
$h$ {\sl just above} $\underline{h}(\gl)$. As a matter of fact
a natural choice is for example  $h=h ^{(0.67)}(\gl)>\underline{h}(\gl)$, recall
\eqref{eq:hm}, for a set of values of $\gl$, and this is part of the content of
Table \ref{tbl:2}: in particular $\bbE \log Z_{2N_+,\go}^{\gl,h^{(0.67)}(\gl)}(0)  >0$
with a probability of error smaller than $10^{-5}$ for the values of $\gl$
between $0.1$ and $1$. However we stress that
for some of these $\gl$'s we have a much smaller $p$--value, see the caption
of Table \ref{tbl:2}, and that the content of this table is much richer
and it approaches also the question of whether or not
a symbolic computation  or some other form of
computer assisted argument could lead to $h_c(\gl)>\underline{h}(\gl)$ for
some $\gl$, and therefore for $\gl$ in an interval. Since such an argument would require
$N$ to be {\sl small}, intuitively the hope resides in large values of $\gl$, recall also
  Remark~\ref{rem:N=2}. It turns out that one needs in any case
  $N$ larger than $700$ in order to observe a localization phenomenon
  at $h^{(0.67)}(\gl)$.
We now give some details on the procedure that leads to Table  \ref{tbl:2}.

\begin{table}[h]
\begin{center}
\begin{tabular}{|c|c|c|c|c|c|c|c|c|c|}
\hline
$\gl $ &$0.05 (\star)$&$0.1$&$0.2$&$0.4$&$0.6$&$1$&$2(\star)$&$4(\star\star )$&$8(\star\star )$\\
\hline
$N_+$&750000&190000&40000&9500&4250&1800&900&800&800\\
\hline
$N_-$&600000&130000&33000&7500&3650&1550&750&700&700\\
\hline
\end{tabular}
\end{center}
\caption{\label{tbl:2}
For a given $\gl$, both $\bbE \log Z_{2N_+,\go}^{\gl,h^{(0.67)}(\gl)}(0)  >0$ and
$\bbE \log Z_{2N_-,\go}^{\gl,h^{(0.67)}(\gl)}(0)  <0$ with a probability of error
smaller than $10^{-5}$ (and in some cases much smaller than that).
Instead for the two cases marked by a $(\star)$ the level of error
is rather between $10^{-2}$ and  $10^{-3}$. For large values of $\gl$, the two cases marked with
 $(\star \star)$, it becomes computationally
expensive to reach small $p$--values. However, above $\gl=3$ one observes
that the values of $Z_{2N,\go}(0)$
essentially do not depend anymore on the value of $\gl$. This can be interpreted
in terms of convergence to a limit ($\gl \to \infty$) model, as it is explained in Remark~\ref{rem:starstar}.
If we then make the hypothesis that this limit model sharply describes the
copolymer along the curve $(\gl, h^{(m)}(\gl))$ for $\gl$ sufficiently large and we apply the concentration
inequality, then the given values of $N_+$  and $N_-$ are tested
with a very small probability of error. Since the details of such a procedure are
quite lengthy we do not report them here.
We have constructed (partial) tables also for different laws of $\go$,
notably $\go_1 \sim N(0,1)$,  and they turned out
to yield larger, at times substantially larger, values of $N_\pm (\gl)$. 
}
\end{table}

First and foremost, the concentration argument
that leads to Proposition~\ref{th:stat} is symmetric
and it works for deviations below the mean as well as above.
So we can, in the very same way, test the null hypothesis
$\bbE\log Z_{2N, \go}(0)> 0$ and, possibly,  refuse it
if $\hat u _n <0$, exactly with the same $p$--value
as in Proposition~\ref{th:stat}.
Of course an important part of Proposition~\ref{th:stat} was coming
from the finite volume localization condition \eqref{eq:loc_char}:
 we do not have an analogous statement for delocalization
(and we do not expect that there exists one).
But, even if $\bbE\log Z_{2N, \go}(0)\le 0$ does not imply delocalization,
it says at least that it is pointless to try to prove localization
by looking at a system of that size.

In Table~\ref{tbl:2} we show
two values of the system size $N$,  $N_+$ and $N_-$,
for which, at a given $\gl$, one has that
$\bbE\log Z_{2N_+, \go}(0)> 0$ and  $\bbE\log Z_{2N_-, \go}(0)< 0$
with a fixed probability of error (specified in the caption of the Table).
It is then reasonable to guess that the transition from
negative to positive values of $\bbE\log Z_{\cdot , \go}(0)$ happens
for $N\in (N_-,N_+)$. There is no reason whatsoever to expect that
 $\bbE\log Z_{N, \go}(0)$ should be monotonic in $N$ but according to our numerical result it is not unreasonable to expect that monotonicity should set in for $N$ large or, at least, that for $N< N_-$ (respectively $N>N_+$)
 $\bbE\log Z_{2N, \go}(0)$ is definitely negative
(respectively positive).

\begin{figure}[h]
\begin{center}
\leavevmode
\epsfysize =8 cm
\psfragscanon
\psfrag{N}[c][l]{$N$}
\psfrag{lambda}[c][l]{ $\gl$}
\epsfbox{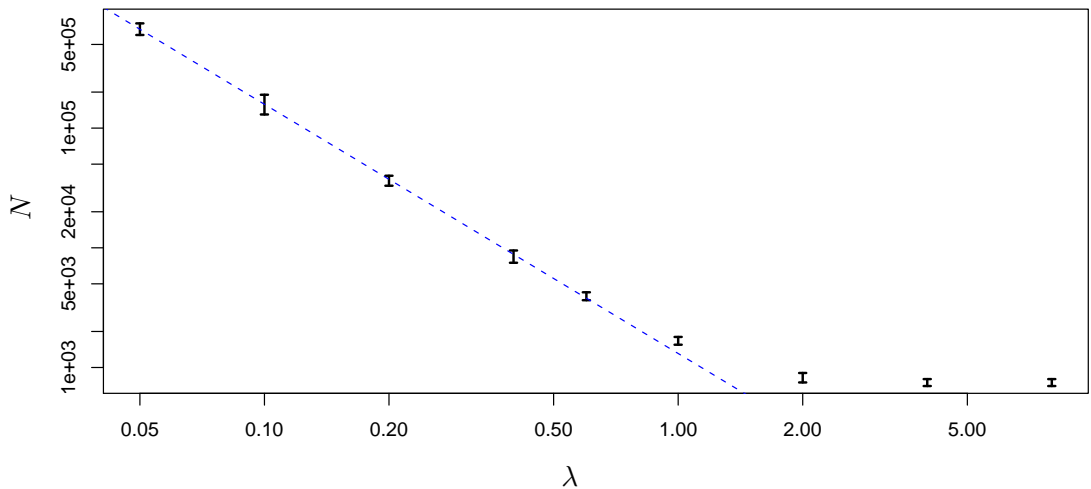}
\end{center}
\caption{\label{fig:NCA} A graphical representation  of
Table~\ref{tbl:2}. The plot is log--log, and a $\gl^{-c}$ behavior is rather
evident, $c$ is about $2.08$. This can be nicely interpreted in terms
of the coarse graining technique in the proof of the weak interaction scaling
limit of the free energy in \cite{cf:BdH}: from that argument one extracts that
if $\gl $ is small the excursions that give a contribution to the free
energy have {\sl typical} length $\gl^{-2}$ and that
in the limit the polymer is just made up by this type of excursions.
One therefore  expects
that it suffices a system of size $N(\gl)$, with $\lim_{\gl \searrow 0} \gl^2 N(\gl)=+\infty$,
to observe localization if $m<h_c^\prime( 0)$, $h=h^{(m)}(\gl)= m \gl (1+o(1))$ and $\gl$ is small.}
\end{figure}

\begin{rem}
\label{rem:starstar}
\rm
As pointed out in the caption of Table  \ref{tbl:2}, from numerics one observes
a very sharp convergence to a $\gl$ independent behavior as $\gl$
becomes large, along the line $h=h^{(m)}(\gl)$.
This is easily interpreted if one observes that $h^{(m)}(\gl)= 1-((\log 2) /2m\gl)+O(\exp(-4m\gl))$
so that
\begin{equation}
\label{eq:lim_mod}
\lim_{\gl \to \infty} \exp\left(-2 \gl \sum_{n=1}^N
\left( \go_n +h\right) \Delta_n \right)\, =\, \exp\left(\frac{\log 2}{m}\sum_{n=1}^N \Delta_n\right)
\ind_{\left\{\sum_{n=1}^N \Delta_n (1+\go_n)=0\right\} } (S).
\end{equation}
This corresponds to the model where a positive charge never enters the lower half-plane and where the energy of a configuration is proportional to the number of negative charges in the lower half-plane.
\end{rem}


\bigskip

\section{Lower bound strategies versus the true strategy}
\label{sec:lb}

\subsection{An approach to lower bounds on the critical curve}
\label{sec:lb_outline}
In this section we give an outline of a new derivation of the lower bound
\begin{equation}
\label{eq:mainBG}
\underline h (\gl)\le h_c(\gl),
\end{equation}
with $\underline h (\gl)$ defined in (\ref{eq:sumupq}). The complete proof may be found
in Appendix \ref{app:prooflb}. The argument takes inspiration from the ideas used in the proof of Proposition 3.1 in \cite{cf:GT} and, even if it is essentially the proof of \cite{cf:BG} in disguise,
in the sense that the selection of the random walk trajectories that are kept
and whose energy contribution is evaluated
does not differ too much (in a word: the {\sl strategy} of the polymer
is similar), it is however conceptually somewhat different and it
will naturally lead to some considerations on the precise
asymptotic behavior of $Z_{N, \go}$ in the delocalized phase and
even in the localized phase close to criticality.

\bigskip

The first step in our proof of (\ref{eq:mainBG}) is a different way of looking at localization. For any fixed positive number $C$ we introduce the stopping time (with respect to the natural filtration of the sequence $\{\go_n\}$) $T^C = T^{C,\gl,h}(\go)$ defined by
\begin{equation}
\label{eq:T^C}
T^{C,\gl,h}(\go) := \inf \{N\in 2\N:\ Z_{N,\go}^{\gl,h}(0) \ge C\}\,.
\end{equation}

The key observation is that if $\bbE[T^C] < \infty$ for some $C>1$, then the polymer is localized. Let us sketch a proof of this fact (for the details, see Proposition~\ref{prop:stopping1}): notice that by the very definition of $T^C$ we have $Z_{T^C(\go),\go}(0) \ge C$. Now the polymer that is in zero at $T^C(\go)$ is equivalent to the original polymer, with a translated environment $\go'=\theta^{T^C} \go$, and setting $T_2(\go) := T^C (\go')$ we easily get $Z_{T_1(\go)+T_2(\go),\go}(0) \ge C^2$ (we have put $T_1(\go):=T^C(\go)$). Notice that the new environment $\go'$ is still typical, since $T^C$ is a stopping time, so that $T_2$ is independent of $T_1$ and has the same law. This procedure can be clearly iterated, yielding an IID sequence $\{T_i(\go)\}_{i=1,2,\ldots}$ that gives the following lower bound on the partition function:
\begin{equation} \label{eq:lowbound}
Z_{T_1(\go)+\ldots +T_n(\go),\go}(0) \ge  C^n\,.
\end{equation}
From this bound one easily obtains that
\begin{equation}
\label{eq:forappB}
\tf(\gl,h) \stackrel{\text{a.s.}}{= } \lim_{n\to\infty} \frac{\log Z_{T_1(\go)+\ldots +T_n(\go),\go}(0)}{T_1(\go)+\ldots +T_n(\go)} \ge \frac{\log C}{\bbE[T^C]}\,,
\end{equation}
where we have applied the strong law of large numbers, and localization follows since by hypothesis $C>1$ and $\bbE[T^C]<\infty$.

\medskip

\begin{rem}
\label{rem:reciprocal}
\rm
It turns out that also the reciprocal of the claim just proved holds true, that is \textsl{the polymer is localized if and only if $\bbE[T^C]<\infty$}, with an arbitrary choice of $C>1$, see Proposition~\ref{prop:stopping1}. In fact the case $\bbE[T^C]=\infty$ may arise in two different ways:
\begin{enumerate}
\item the variable $T^C$ is defective, $\bbP[T^C=\infty]>0$: in this case with positive probability $\{Z_{N,\go}(0)\}_N$ is a bounded sequence, and delocalization follows immediately;
\item\label{en:scenario} the variable $T^C$ is proper with infinite mean, $\bbP[T^C=\infty]=0,\ \bbE[T^C]=\infty$: in this case we can still build a sequence $\{T_i(\go)\}_{i=1,2,\ldots}$ defined as above and this time the lower bound \eqref{eq:lowbound} has \textsl{subexponential} growth. Moreover it can be shown that in this case the lower bound \eqref{eq:lowbound} gives the true free energy, cf. Lemma~\ref{lem:chop}, which therefore is zero, so that delocalization follows also in this case.
\end{enumerate}
As a matter of fact, it is highly probable that in the interior of the delocalized phase
$Z_{N,\go}(0)$ vanishes $\bbP (\dd \go)$--a.s.
when $N \to \infty$ and this would rule out the scenario (\ref{en:scenario}) above, saying that for $C>1$ the random variable $T^C$ must be either integrable or defective. We take up again this point in Sections~\ref{sec:path} and~\ref{sec:guess}: we feel that this issue
is quite crucial in order to fully understand the delocalized phase
of disordered models.
\end{rem}


\begin{rem}
\label{rem:analogy}
\rm
Dealing directly with $T^C$ may be difficult.
Notice however that if one finds
 a random time (by this we mean simply an integer--valued random variable)
 $T=T(\go)$ such that
 \begin{equation} \label{eq:but}
Z_{T(\go), \go}(0)\ge C>1\,, \qquad \text{with \ } \bbE[T]<\infty\,,
\end{equation}
then localization follows. This is simply because
this implies $T^C \le T$ and hence $\bbE[T^C]<\infty$. Therefore localization is equivalent to the condition $\log Z_{T(\go), \go}(0)>0$ for an \textsl{integrable} random time $T(\go)$: we would like to stress the analogy between this and the criterion for localization given in \S~\ref{sec:superadd}, see \eqref{eq:loc_char}.
\end{rem}

\medskip

Now we can turn to the core of our proof: we are going to show that for every $(\gl,h)$ with $h<\underline h (\gl)$ we can build a random time $T=T(\go)$ that satisfies \eqref{eq:but}.
The construction of $T$
 is based on the  idea that for $h>0$ if localization prevails
 is because of rare $\go$--stretches that invite the polymer
 to spend time in the lower half--plane in spite of the action of $h$.

\begin{figure}[h]
\begin{center}
\leavevmode
\epsfysize =5 cm
\psfragscanon
\psfrag{0}[c][l]{$0$}
\psfrag{l}[c][l]{ $\ell$}
\psfrag{n}[c][l]{ $n$}
\psfrag{h}[l][l]{ $L$}
\psfrag{S}[c][l]{$S$}
\psfrag{T}[c][l]{$T(\go)$}
\epsfbox{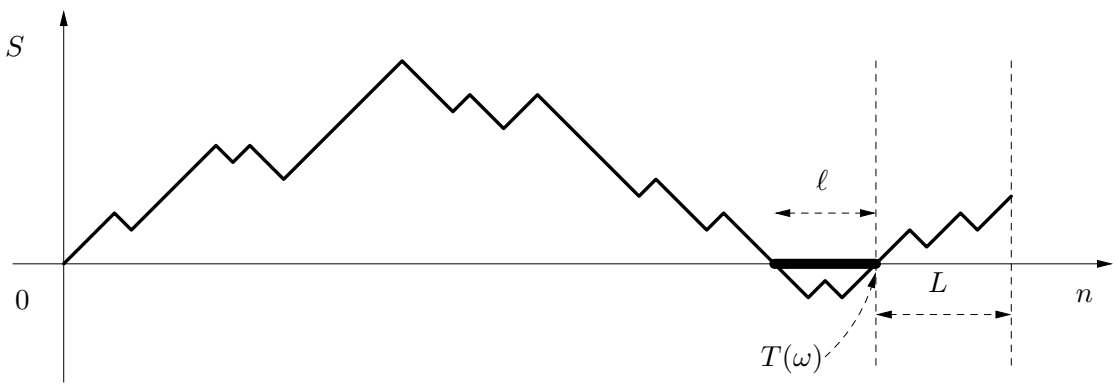}
\end{center}
\caption{\label{fig:ZT}
Inequality \eqref{eq:beyondT}
comes simply from restricting the evaluation of $Z_{T(\go) +L, \go} $ to the trajectories
visiting the {\sl $q$--atypical} stretch of length $\ell$  and by staying away from
the unfavorable solvent after that.
}
\end{figure}

The strategy we use consists in looking
for $q$--atypical stretches of length at least $M\in 2\N$, where $q<-h$ is
the average charge of the stretch. Rephrased a bit more precisely,
we are looking for the smallest $n\in 2\N$ such that
$\sum_{i=n-k+1}^n \go_i /k <q$ for some even integer $k \ge M$.
It is well known that such a random variable grows, in the sense
of Laplace, as $\exp(\Sigma(q) M)$ for $M\to \infty$, where $\gS(q)$ is the Cramer functional
\begin{equation}
\label{eq:cramer}
\gS(q) := \sup_{\ga\in\R} \{\ga q-\log \M(\ga)\}\,.
\end{equation}
One can also show without much effort that the length of such a stretch
cannot be much longer than $M$.
Otherwise stated, this is the familiar statement that the longest $q$--atypical
sub--stretch of $\go_1, \ldots , \go_N$ is of typical length $\sim \log N/\Sigma (q)$.
 So $T(\go)$ is for us the end--point
of a $q$--atypical stretch of length approximately $(\log T(\go))/\Sigma(q)$:
by looking for sufficiently long $q$--atypical stretches
we have always the freedom to choose $T(\go) \gg 1$, in such a
way that also $\log T(\go) \ll T(\go)$ and this is helpful for the estimates.
So let us bound $Z_{T(\go), \go}$ from
below by considering only the trajectories of the walk that
stay in the upper half--plane up to the beginning of the $q$--atypical stretch and
that are negative in the stretch, coming back to zero at step $T(\go)$
(see Fig.~\ref{fig:ZT}: the polymer is cut at the first dashed vertical line).
The contribution of these trajectories is easily evaluated: it is approximately
\begin{equation}
\label{eq:int1}
\left(
\frac {1} {T(\go)^{3/2}}
\right)
 \exp\left( -2\gl (q+h)  \frac{\log T(\go)}{\Sigma(q)}\right).
\end{equation}
For such an estimate we have used
\eqref{eq:asympt} and $\log T(\go) \ll T(\go)$
both in writing the probability that the first return to zero
of the walk is at the beginning of the $q$--atypical stretch and in neglecting the probability
that the walk is negative inside the stretch.
It is straightforward to see that if
\begin{equation}
\label{eq:int2}
\frac {4\gl} 3 h < -\frac {4\gl} 3 q -  \Sigma(q),
 \end{equation}
and if $T(\go)$ is large, then also  the quantity in
 \eqref{eq:int1} is large. We can still optimize this procedure by choosing $q$ (which must be sufficently negative, i.e. $q < -h$).
 By playing with
\eqref{eq:cramer}
one sees that one can choose $q_0\in \R$ such that for $q=q_0$ the right--hand side
in \eqref{eq:int2} equals $\log \M (-4\gl/3)$ and
if $h<\log \M (-4\gl/3)/(4\gl/3) = \underline{h} (\gl)$
 then $q_0 <-h$. This argument therefore is saying that there
 exists $C>1$ such that
 \begin{equation}
 \label{eq:C1}
 Z_{T(\go), \go}(0) \, \ge \, C,
\end{equation}
for every $\go$. It only remains to show that $\bbE[T]<\infty$: this fact, together with a detailed proof of the argument just presented can be found in Appendix~\ref{app:prooflb}.

\medskip

\begin{figure}[h]
\begin{center}
\leavevmode
\epsfysize =16 cm
\psfragscanon
\psfrag{x}[c][l]{\small $N$}
\psfrag{y}[c][l]{\small $\log  Z_{2N, \go}^{\gl, h}$}
\psfrag{a}[l][l]{A: $h=0.42$}
\psfrag{b}[l][l]{B:  $h=0.44$}
\psfrag{c}[l][l]{C:  $h=0.43$}
\psfrag{d}[l][l]{D:  $h=0.43$ (Zoom)}
\epsfbox{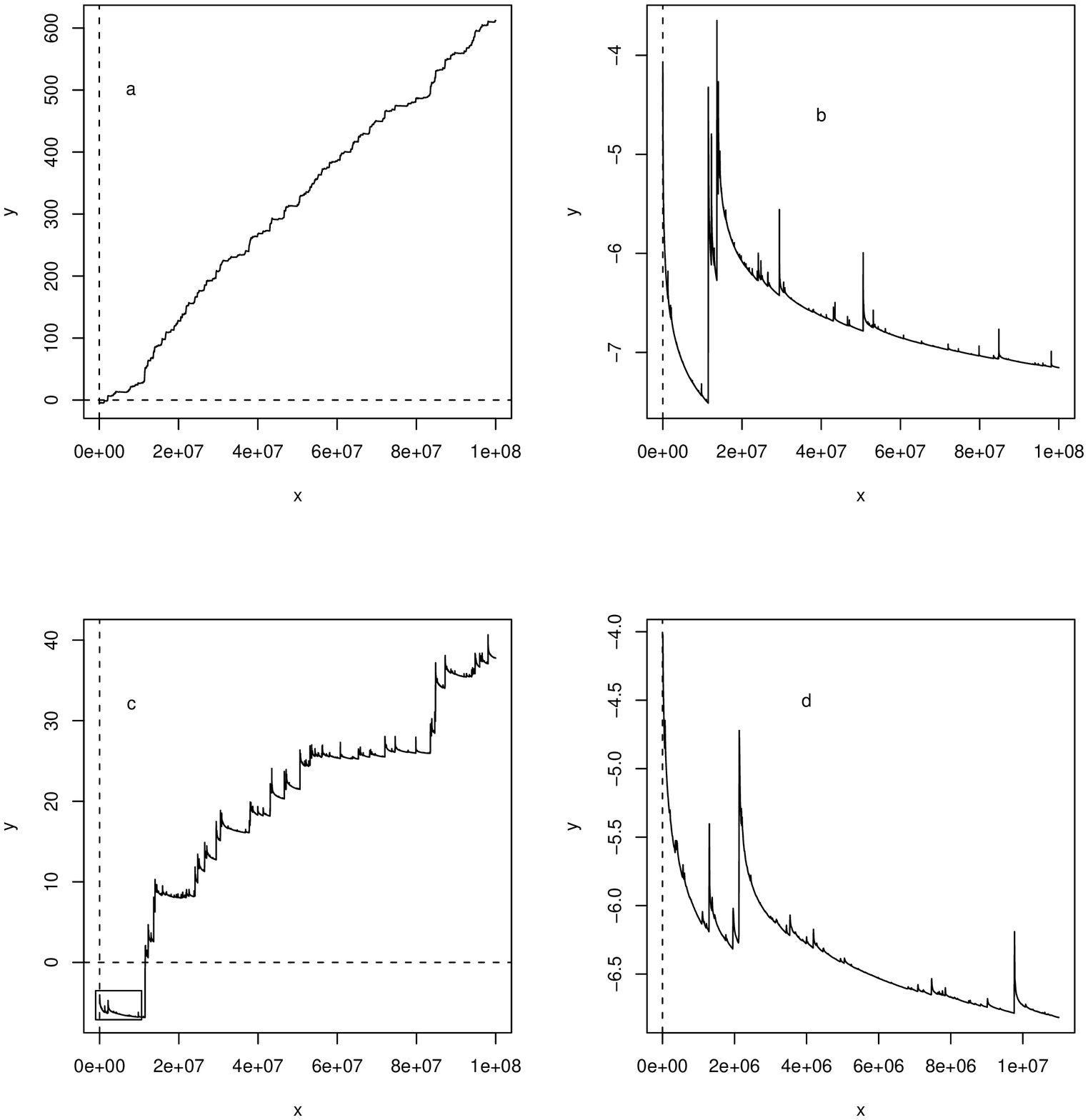}
\end{center}
\caption{
\label{fig:tmp}
For $\gl=0.6$ ($\underline h (0.6)\simeq 0.36$ and $\overline h (0.6)\simeq 0.49$) , the behavior of
$\log Z_{2N, \go}$ for $h=0.42$ (A), $0.43$ (C,D) and $0.44$ (B). The sequence
of charges is the same in all the cases.
In case A, the polymer is localized with free energy approximately
$3\cdot 10^{-6}$: the linear growth is quite clear, but a closer look shows sudden jumps,
which correspond to atypically negative stretches of charges.
Getting closer to the critical point, case C, the growth is still rather evident, but
it is clearly the result of sudden growths followed by slow decays (approximately polynomial
with exponent $-1/2$).  Case B suggests delocalization: a closer analysis reveals a decay
of the type $N ^{-1/2}$, but sharp deviations are clearly visible and these deviations
are in reality much larger, since in the graph we have plotted just one point every 10000.
Case D is the zoom of the rectangle in the left corner of C.
The similarity between B and D
 make clear that claiming delocalization looking at the behaviour of the partition function is difficult.
}
\end{figure}

\subsection{Persistence of the effect of rare stretches}
As pointed out in the previous section, there is strong evidence
that $h_c(\gl)> \underline{h} (\gl)$.
At this stage Fig.~\ref{fig:tmp}  is of particular interest.
Notice first of all that in spite of being substantially above $\underline{h}(\cdot)$
the copolymer appears to be still localized, see in particular case A.

The rigorous lower bounds that we are able
to prove cannot establish localization in the region we are considering.
All the same, notice that if one does  not cut the polymer at $T(\go)$,
as in the argument above, but at $T(\go)+L$,
a  lower bound of the following type
\begin{equation}
\label{eq:beyondT}
Z_{T(\go)+L, \go}\, \stackrel{\text{roughly}}{\ge}\,
\text{const.} \frac1{T(\go)^{3/2}}\,
\exp\left( -2\gl (q+h)\frac{\log T(\go)}{\Sigma (q)}\right) \, \frac 1 {L^{1/2}},
\end{equation}
is easily established. Of course we are being imprecise, but we just want to convey
the idea, see also Fig.~\ref{fig:ZT}, that after passing through an atypically {\sl negative}
stretch of environment ($q>0$), the effect of this stretch decays at most like $L^{-1/2}$,
that is the probability that a walk stays  positive for a time $L$.

At this point we stress that the argument outlined in \S~\ref{sec:lb_outline}
and re--used for \eqref{eq:beyondT}
may be very well applied to $h> \underline{h}(\gl)$, except that
this time it does not suffice for $\eqref{eq:C1}$.
But it yields nevertheless  that for $h \in \left(\underline{h}(\gl), \overline{h}(\gl)\right)$
the statement $Z_{N,\go}\sim N^{-1/2}$, something a priori expected (for
example \cite{cf:BH})  
in the delocalized regime and true for non disordered systems, is
violated. More precisely, one can find a sequence of random
times $\{\tau_j\}_j$, $\lim_j \tau _j= \infty$ such that
$Z_{\tau_j,\go}\ge {\tau_j}^{-1/2+a}$, $a=a(\gl,h)>0$ (see  Proposition~4.1 in \cite{cf:GT}).
These random times are constructed exactly by looking
for $q$--atypical stretches as above and one can appreciate
such an irregular decay for example in case B of Fig.~\ref{fig:tmp},
and this in spite of the fact that the data have been  strongly coarse grained.

\smallskip

Therefore the lower bound
\eqref{eq:beyondT}, both in the localized and in the delocalized regime,
yields the following  picture:
the lower bound we found on  $Z_{N, \go}$ grows suddenly in correspondence of atypical stretches
and after that it decays with an exponent $1/2$, up to another atypical stretch.
This matches Fig.~\ref{fig:tmp}, at least on a qualitative level, see the caption
of the figure.


Of course it very natural to ask what is missing, on a theoretical
level, to the strategy that we are adopting for the lower bound to
match the quantitative discrepancy. Moreover,
since the $\go$ sequence is of course known, 
one may look at the atypical stretches, this time
defined by the points of sudden growth of  $Z_{N, \go}$,
and look for the specificity of such stretches.
Up to now we have not been able to extract from 
this analysis definite answers.




\section{The delocalized phase: a path analysis}
\label{sec:path}

Let us start with a qualitative observation: if we set the parameters $(\gl,h)$ of the copolymer to $(\gl, h^{(m)}(\gl))$ with $m = 0.9$, then the observed behavior of $\{Z_{N,\go}^{\gl,h}(0)\}_N$ --suitably averaged over blocks in order to eliminate local fluctuations-- is somewhat close to $\text{(const)}/N^{3/2}$. This is true for all the numerically accessible values of $N$ (up to $N\sim 10^8$), at once for a number of values of~$\gl$ and for a great number of typical environments $\go$. Of course this is suggesting that for $m=0.9$ the curve $h^{(m)}(\gl)$ lies in the delocalized region, but it is not easy to convert this qualitative observation into a precise statement, because we do not have a rigorous finite--volume criterion to state that a point $(\gl,h)$ belongs to the delocalized phase (the contrast with the localized phase, see~\eqref{eq:loc_char}, is evident). In other words, we cannot exclude the possibility that the system is still localized but with a characteristic size much larger than the one we are observing.

Nevertheless, the aim of this section is to give an empirical criterion, based on an analysis of the path behavior of the copolymer, that will allow us to provide some more quantitative argument in favor of the fact that the curve $h^{(m)}(\gl)$ lies in the delocalized region even for values of~$m<1$. This of course would entail that the upper bound $\overline h(\gl)$ defined in \eqref{eq:sumupq} is not strict.

\subsection{Known and expected path behavior}
\label{sec:path1}
We want to look at the whole \textsl{profile} $\{Z_{N,\go^r}^{\gl,h}(x)\}_{x\in\Z}$ rather than only at $Z_{N,\go^r}^{\gl,h}(0)$, where by $\go^r$ we mean the environment $\go$ in the {\sl
 backward direction}, that is $(\go^r)_n := \go_{N+1-n}$ (the reason for this choice is explained in Remark~\ref{rem:backward} below). The link with the path behavior of the copolymer, namely the law of $S_N$ under the polymer measure $\bP_{N,\go^r}^{\gl,h}$, is given by
\begin{equation}
    \frac{Z_{N,\go^r}^{\gl,h}(x)}{Z_{N,\go^r}^{\gl,h}} = \bP_{N,\go^r}^{\gl,h} (S_N = x)\,.
\end{equation}

As already remarked in the introduction, although the localized and delocalized phases have been defined in terms of free energy they do correspond to sharply different path behaviors. In the localized phase it is known \cite{cf:Sinai,cf:BisdH} that the laws of $S_N$ under $\bP_{N,\go^r}^{\gl,h}$ are \textsl{tight}, which means that the polymer is essentially at $O(1)$ distance from the $x$--axis. The situation is completely different in the (interior of the) delocalized phase, where one expects that $S_N = O(\sqrt{N})$: in fact the conjectured path behavior (motivated by the analogy with the known results for non disordered models, see in particular \cite{cf:MGO},  \cite{cf:DGZ} and \cite{cf:CGZ}) should be weak convergence under diffusive scaling to the \textsl{Brownian meander process} (that is Brownian motion conditioned to stay positive on the interval $[0,1]$, see \cite{cf:RevYor}). Therefore in the (interior of the) delocalized phase the law of $S_N/\sqrt{N}$ under $\bP_{N,\go^r}^{\gl,h}$ should converge weakly to the corresponding marginal of the Brownian meander, whose law has density $x \exp(-x^2/2) \ind_{(x\ge 0)}$.

We stress however that for the delocalized regime the rigorous results that are available are more meager: essentially the only known $\bbP(\dd\go)$--a.s. result is that for any $L>0$
\begin{equation}
    \lim_{N\to\infty} \bE_{N,\go^r}^{\gl,h} \Bigg[ \frac 1N \sum_{n=1}^N \ind_{(S_n \ge L)} \Bigg] = 1 \qquad \bbP(\dd\go)\text{--a.s.}\,,
\end{equation}
that is the polymer spends almost all the time above any prefixed level. More precise results have been derived for the path behavior of the polymer under the \textsl{quenched averaged measure} $\bbE \bE_{N,\go}^{\gl,h} [\,\cdot\, ]$\,: these results go in the direction of proving the conjectured scaling limit, but they still do not suffice (we refer to \cite{cf:GT} for more details and also for a discussion on what is still missing).

In spite of the lack of precise rigorous results,
the analysis we are going to describe is carried out under the hypothesis that, in the interior of the delocalized phase, the scaling limit towards Brownian meander holds true (as it will be seen, the numerical results provide a sort of \textsl{a posteriori} confirmation of this hypothesis).

\begin{rem} \label{rem:backward} \rm
From a certain point of view attaching the environment backwards does not change too much the model: for example it is easy to check that if one replaces $\go$ by $\go^r$ in \eqref{eq:free_energy}, the limit still exists $\bbP(\dd\go)$--a.s. and in $\bbL_1(\dd\bbP)$. Therefore the free energy is the same, because $\{\go^r_n\}_{1\le n \le N}$ has the same law as~$\{\go_n\}_{1\le n \le N}$, for any fixed~$N$.

However, if one focuses  on the law of $S_N$ as a function of~$N$ \textsl{for a fixed environment} $\go$, the behavior reveals to be much smoother under $\bP_{N,\go^r}^{\gl,h}$ than under $\bP_{N,\go}^{\gl,h}$. For instance, under the original polymer measure $\bP_{N,\go}^{\gl,h}$ it is no more true that in the localized region the laws of $S_N$ are tight (it is true only {\sl most of the time}, see \cite{cf:G} for details). The reason for this fact is to be sought in the presence of long {\sl atypical} stretches in every typical~$\go$ (this fact has been somewhat quantified in \cite[Section 4]{cf:GT} and it is at the heart of the approach in Section~\ref{sec:lb}) that are encountered along the copolymer as $N$ becomes larger. Of course the effect of these stretches is very much damped with the backward environment.

A similar and opposite phenomenon takes place also in the delocalized phase. In fancier words, we could say that for fixed $\go$ and as $N$ increases, the way $S_N$ approaches its {\sl limiting behavior} is faster when the environment is attached backwards: it is for this reason that we have
chosen to work with $\bP_{N,\go^r}^{\gl,h}$.
\end{rem}

\subsection{Observed path behavior: a numerical analysis}
In view of the above considerations, we choose as a measure of the delocalization of the polymer the
$\ell_1$ distance $\bigtriangleup _N^{\gl,h}(\go)$ between the numerically computed profile for a polymer of size $2N$ under $\bP_{2N,\go^r}^{\gl,h}$, and the conjectured asymptotic delocalized profile:
\begin{equation}
    \bigtriangleup _N^{\gl, h}(\go) := \sum_{x \in 2\Z} \Bigg| \frac{Z_{N,\go^r}^{\gl,h}(x)}{Z_{N,\go^r}^{\gl,h}} \;-\; \frac{1}{\sqrt{2N}} \, \varphi^+ \bigg( \frac{x}{\sqrt{2N}} \bigg) \Bigg|\,,  \qquad \gp^+(x) := x \, e^{-x^2/2} \ind_{(x\ge 0)}\,.
\end{equation}
Loosely speaking, when the parameters $(\gl,h)$ are in the interior of the the delocalized region we expect $\bigtriangleup _N$ to decrease to~$0$ as $N$ increases, while this certainly will not happen if we are in the localized phase.

\smallskip

The analysis has been carried out at $\gl=0.6$: we recall that the lower and upper bound of \eqref{eq:sumupq} give respectively $\underline h(0.6) \simeq 0.36$ and $\overline h(0.6) \simeq 0.49$, while the lower bound we derived with our  test for localization is $h=0.41$, see Table~\ref{tbl:1}. However,
as observed in Section \ref{sec:lb}, Fig.~\ref{fig:tmp}, there is numerical evidence that $h=0.43$ is still localized, and for this reason  we have analyzed the values of $h=0.44, 0.45, 0.46, 0.47$ (see below for an analysis on smaller values of~$h$).

\begin{figure}[h]
\begin{center}
\leavevmode
\epsfysize =7.9 cm
\epsfbox{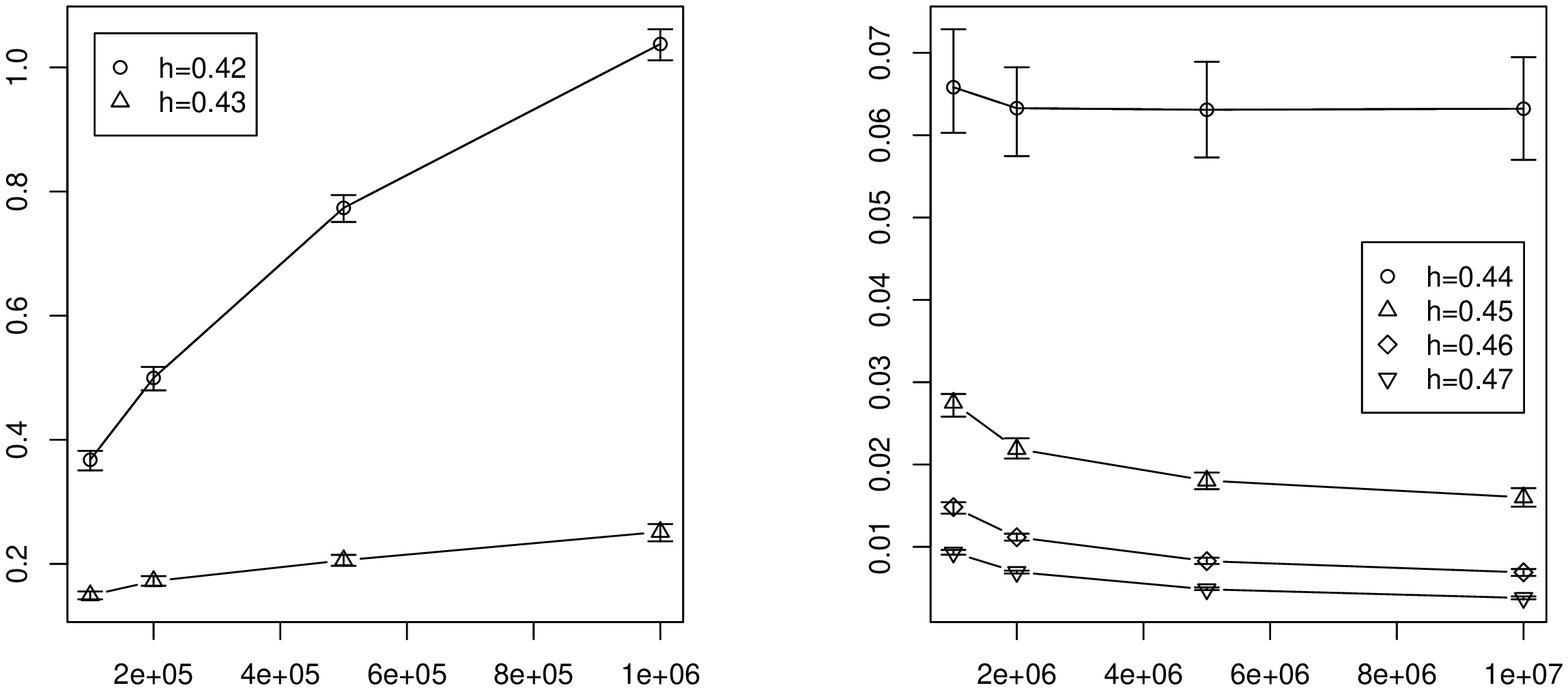}
\end{center}
\caption{
\label{fig:F}
Graphical representation of the data of Tables~\ref{tbl:F} (on the right) and~\ref{tbl:F2} (on the left). The plotted points are the sample medians against the sample size, the error bars correspond to the confidence intervals given in Tables~\ref{tbl:F} and~\ref{tbl:F2}.
}
\end{figure}

For each couple $(\gl, h)$ we have computed $\bigtriangleup _{N}^{\gl,h}(\go)$ for the sizes $N=a\times 10^6$ with $a=1,2,5,10$ and for $500$ independent environments. Of course some type of statistical analysis must be performed on the data in order to decide whether there is a decay of~$\bigtriangleup $ or not. The most direct strategy would be to look at the sample mean of
a family of IID variables distributed like $\bigtriangleup _N(\go)$, but it turns out that the fluctuations are too big to get reasonable confidence intervals for this quantity (in other words, the sample variance does not decrease fast enough), at least for the numerically accessible sample sizes. A more careful analysis shows that the variance is essentially due to a \textsl{very small} fraction of data that have \textsl{large} deviations from the mean, while the most of the data mass is quite concentrated.

\smallskip
\begin{rem}\rm
It is actually interesting to observe that the rare samples
that {\sl affect} the sample  variances are in reality very close to meanders
anyway, only with a smaller variance. This is the signature of the presence
of atypical pinning stretches in the $\go$--sequence close to the boundary.
A fine analysis of this aspect would lead us too far and it is left for 
future investigation.
\end{rem}
\smallskip

We have therefore chosen 
to focus on the \textsl{sample median} rather than on the sample mean. Table~\ref{tbl:F} contains the results of the analysis (see also Fig.~\ref{fig:F} for a graphical representation): for each value of $h$ we have reported the standard $95\%$ confidence interval for the sample median (see Remark~\ref{rem:conf_int} below for details) for the four different values of~$N$ analyzed. While for $h=0.44$ the situation is not clear, we see that for the values of $h$ greater than $0.45$ there are quantitative evidences for a decrease in~$\bigtriangleup_{N}$: this leads us to the conjecture that the points $(\gl,h)$ with $\gl=0.6$ and $h\ge 0.45$ (equivalently, the points $(\gl,h^{(m)}(\gl))$ with $m \gtrsim 0.876$) lie in the delocalized region.

\begin{table}[h]
\begin{center}
\begin{tabular}{|c||c|c|c|c|}
\hline
$h \backslash N(\times 10^6) $ & 1 & 2 & 5 & 10\\
\hline
\hline
0.44 & [.0603, .0729] & [.0574, .0682] & [.0572, .0689] & [.0570, .0695] \\
\hline
0.45 & [.0258, .0286] & [.0207, .0232] & [.0170, .0190] & [.0149, .0171] \\
\hline
0.46 & [.0140, .0154] & [.0108, .0116] & [.00792, .00869] & [.00647, .00731] \\
\hline
0.47 & [.00905, .00963] & [.00676, .00711] & [.00475, .00508] & [.00364, .00398] \\
\hline
\end{tabular}
\end{center}
\caption{ \label{tbl:F}
The table contains the standard $95\%$ confidence interval for the median of a sample $\{\bigtriangleup _N^{\gl,h}(\go)\}_{\go}$ of size 500, where $\gl=0.6$ and $h,N$ take the different values reported in the table. For the values of $h \ge 0.45$ the decreasing behavior of $\bigtriangleup_{N}$ is quite evident (the confidence intervals do not overlap), see also Fig.~\ref{fig:F}.}
\end{table}

As already remarked, these numerical observations cannot rule out the possibility that the system is indeed localized, but the system size is too small to see it. For instance, we have seen that there are evidences for $h=0.43$ to be localized (see case~C of Fig.~\ref{fig:tmp}).
In any case, the exponential increasing of $Z_N(0)$ is detectable only at sizes of order$\sim 10^8$, while for smaller system sizes (up to$\sim 10^7$) the qualitative observed behavior of $Z_N(0)$ is rather closer to $(const)/N^{3/2}$, thus apparently suggesting delocalization (see case~D of Fig.~\ref{fig:tmp}).

For this reason it is interesting to look at $\bigtriangleup_N^{0.6,\,h}$ for $h=0.42, 0.43$ and for $N \ll 10^8$. For definiteness we have chosen $N=a\times 10^6$ with $a=1,2,5,10$, performing the computations for $3000$ independent environments: the results are reported in Table~\ref{tbl:F2} (see also Fig.~\ref{fig:F}). As one can see, this time there are clear evidences for an \textsl{increasing} behavior of $\bigtriangleup_N$. On the one hand this fact gives some more confidence on the data of Table~\ref{tbl:F}, on the other hand it suggests that looking at $\{\bigtriangleup_N\}_N$ is a more reliable criterion for detecting (de)localization than looking at $\{Z_N(0)\}_N$.

\begin{table}[h]
\begin{center}
\begin{tabular}{|c||c|c|c|c|}
\hline
$h \backslash N(\times 10^5) $ & 1 & 2 & 5 & 10\\
\hline
\hline
0.42 & [.351, 0.382] & [.480, 0.517] & [.751, 0.794] & [1.01, 1.06] \\
\hline
0.43 & [.143, 0.155] & [.165, 0.180] & [.197, 0.215] & [.236, 0.264] \\
\hline
\end{tabular}
\end{center}
\caption{ \label{tbl:F2}
The table contains the standard $95\%$ confidence interval for the median of a sample $\{\bigtriangleup _N^{\gl,h}(\go)\}_{\go}$ of size 3000, where $\gl=0.6$ and $h,N$ take the values reported in the table. For both values of~$h$ an increasing behavior of $\bigtriangleup_{N}$ clearly emerges, see also Fig.~\ref{fig:F} for a graphical representation.}
\end{table}

\begin{rem} \label{rem:conf_int} \rm
A confidence interval for the sample median can be obtained in the following general way
(the steps below are performed under the assumption that the median
is unique, which is, strictly speaking, not true in our case, but it will be clear that
a finer analysis would not change the outcome). Let $\{Y_k\}_{1\le k \le n}$ denote a sample of size~$n$, that is the variables $\{Y_k\}_k$ are independent with a common distribution, whose median we denote by~$\xi_{1/2}$:
$\bP \left(Y_1 \le \xi_{1/2}\right)=1/2$. Then the variable
\begin{equation}
    \cN_n := \# \{i \le n:\ Y_i \le \xi_{1/2}\}
\end{equation}
has a binomial distribution $\cN_n \sim B(n,1/2)$ and when $n$ is large (for us it will be at least~500) we can approximate $\cN_n/n \approx 1/2 + Z/(2\sqrt{n})$, where $Z \sim N(0,1)$ is a standard gaussian. Let us denote the sample quantiles by $\Xi_q$, defined for $q \in (0,1)$ by
\begin{equation}
    \# \{i \le n:\ Y_i \le \Xi_q\} = \lfloor qn \rfloor\,.
\end{equation}
If we set $a:= |\Phi^{-1}(0.025)|$ ($\Phi$ being the standard gaussian distribution function) then the random interval
\begin{equation}
    \Big[\Xi_{\frac{1}{2}-\frac{a}{2\sqrt{n}}},\; \Xi_{\frac{1}{2}+\frac{a}{2\sqrt{n}}}\Big]
\end{equation}
is a $95\%$ confidence interval for $\xi_{1/2}$, indeed
\begin{align}
    0.95 &= \bP \big( Z \in [-a,a] \,\big) = \bP \bigg( \frac 12 + \frac{1}{2\sqrt{n}}Z \;\in\; \Big[\frac{1}{2}-\frac{a}{2\sqrt{n}}\;,\; \frac{1}{2}+\frac{a}{2\sqrt{n}} \Big] \bigg) \nonumber \\
    &\approx \bP \bigg ( \frac{\cN_n}{n} \in \Big[\frac{1}{2}-\frac{a}{2\sqrt{n}}\;,\; \frac{1}{2}+\frac{a}{2\sqrt{n}} \Big] \bigg) = \bP \bigg( \Xi_{\frac{1}{2}-\frac{a}{2\sqrt{n}}} \le \xi_{1/2} \le \Xi_{\frac{1}{2}+\frac{a}{2\sqrt{n}}} \bigg)\,.
\end{align}
\end{rem}


\bigskip
\section{An empirical observation on the critical curve}
\label{sec:guess}

The key point of this section is that, from a numerical viewpoint,
$h_c(\cdot)$ seems very close to $h^{(m)}(\cdot)$, for a suitable
value of $m$. Of course any kind of
 statement in this direction
 requires first of all a procedure to estimate $h_c(\cdot)$ and
we explain this first.

Our analysis is based on the following
conjecture:
\begin{equation}
\label{eq:conject_h}
(\gl, h)\in \overset{\circ}{\cD} \, \, \Longrightarrow \ \
\lim_{N \to \infty }Z_{2N, \go}^{\gl, h} (0) \, =\, 0, \ \bbP \left( \dd \go \right)-\text{a.s.}.
\end{equation}
The
 arguments in Section~\ref{sec:lb} (and in the Appendix)
suggest the validity of such a conjecture, which  is  comforted  by the numerical observation.
Since, if $(\gl, h)\in \cL$,  $Z_{2N, \go}^{\gl, h} (0)$ diverges (exponentially fast)
$\bbP \left( \dd \go \right)$--almost surely and since $Z_{2N, \go}^{\gl, h} (0)$
is decreasing  in $h$, we define $\hat h _{N, \go} (\gl)$ as the only $h$ that solves
$ Z_{2N, \go}^{\gl, h} (0)=1$.
We expect that $\hat h _{N, \go} (\gl)$ converges to $h_c(\gl)$
as $N$ tends to infinity, for typical $\go$'s.
Of course setting the threshold to the value $1$ is rather arbitrary, but it is
somewhat suggested by \eqref{eq:loc_char} and by  the idea behind
the proof of \eqref{eq:mainBG} (Proposition  \ref{prop:stopping1} and
equation
\eqref{eq:T^C}).

\begin{figure}[h]
\begin{center}
\leavevmode
\epsfysize =8.5 cm
\psfragscanon
\psfrag{x}[c][l]{$\gl$}
\psfrag{xg}[c][l]{$\gl$}
\psfrag{y}[c][l]{$\hat h _{N,\go}(\gl)$}
\psfrag{yg}[c][l]{$\hat h _{N,\go}(\gl)$}
\epsfbox{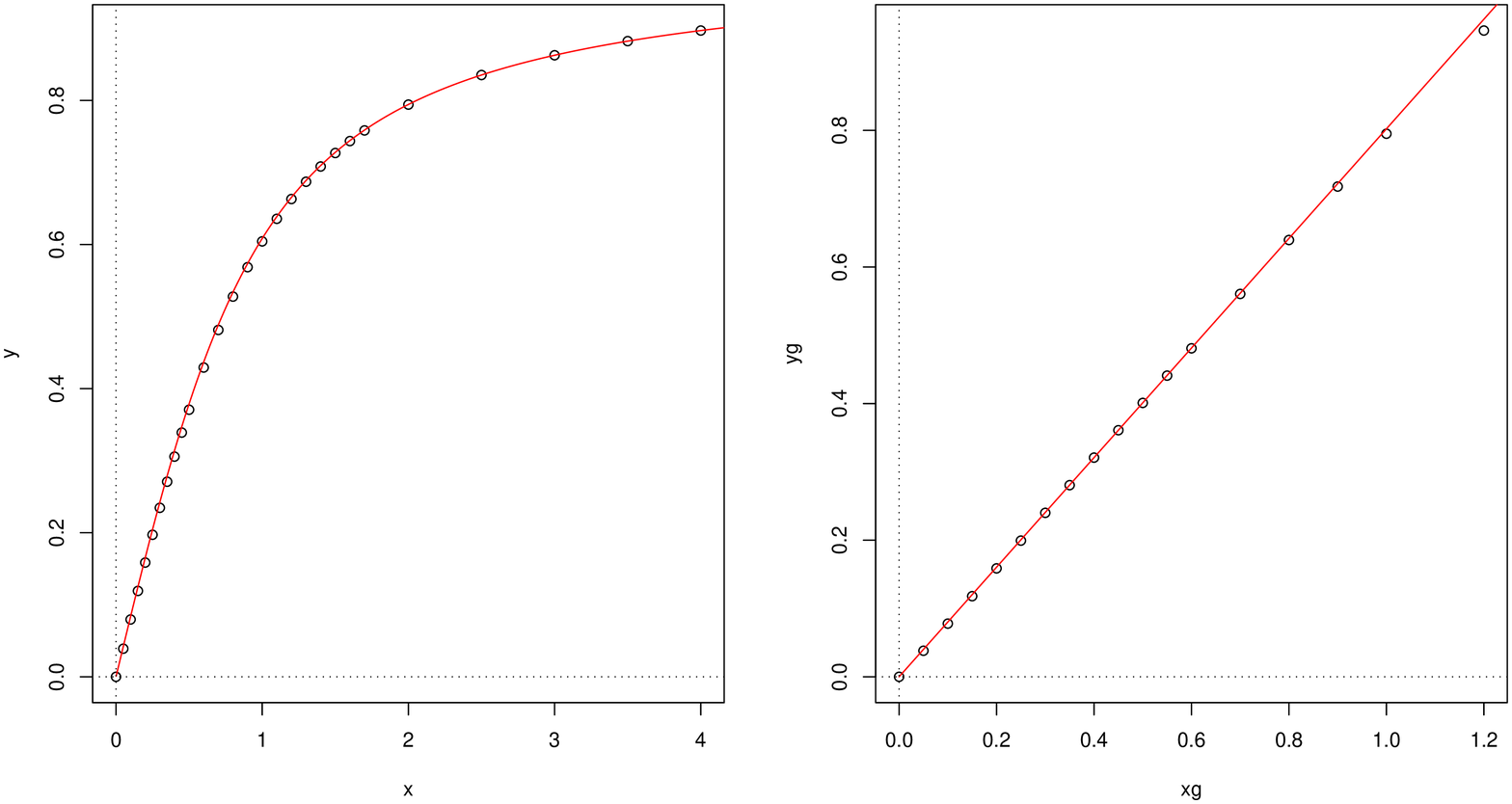}
\end{center}
\caption{
\label{fig:sec5_1}
On the left the case of binary symmetric $\go_1$ and on the right
the case of $\go_1 \sim N(0,1)$, boths for $N= 3.2\cdot 10^7$.
The small circles represent the computed values: the errors on $\hat h _{N,\go}(\gl)$
are
negligible and the plotted points are at the centers of the circles. The continuous
line is instead the curve $h^{(m)}(\cdot)$. In the binary case $m=0.841$ and it has been chosen
by solving $h^{(m)}(4)=\hat h _{N,\go}(4)$. In the Gaussian case $m=0.802$, the maximum
of $ \hat h _{N,\go}(\gl)/\gl$ for the plotted values of $\gl(>0)$.
The rather different values of $\hat m_{N, \go}$ may be somewhat understood
both by considering that these two curves have been obtained for a fixed
realization of $\go$ and by taking into account the remark at the end of the caption
of Table~\ref{tbl:2}: it appears that for  Gaussian charges one needs longer systems in order to get closer
to the values of $m$ observed in the binary case (in particular: for the prolongation,
with the same random
number generator,
of the Gaussian $\go$ sample used here up to $N=5\cdot 10^{7}$ one
obtains $\hat m_{N, \go}=0.812$).
}
\end{figure}

\begin{figure}[h]
\begin{center}
\leavevmode
\epsfysize =7.1 cm
\psfragscanon
\psfrag{Binary}[c][l]{\small $\go_1 =\pm 1, \, \go_1 \sim -\go_1 $}
\psfrag{Gauss}[c][l]{$\go_1 \, \sim \, N(0,1)$}
\psfrag{x}[c][l]{$\gl$}
\psfrag{xg}[c][l]{$\gl$}
\psfrag{y}[c][l]{$\hat h _{N,\go}(\gl)$}
\psfrag{yg}[c][l]{$\hat h _{N,\go}(\gl)$}
\psfrag{tm1}[c][l]{$r_{N,\go}(\gl)$}
\psfrag{tmp1}[c][l]{$r_{N,\go}(\gl)$}
\epsfbox{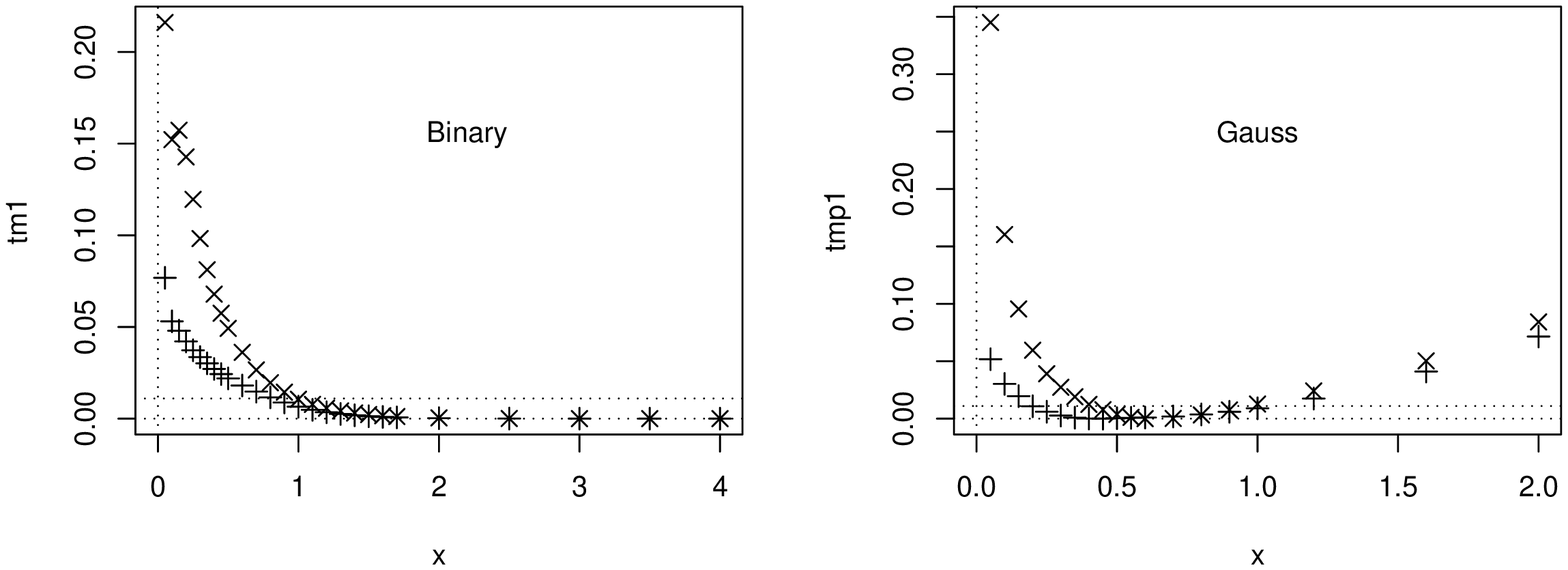}
\end{center}
\caption{
\label{fig:sec5_2}
Relative errors $r_{N,\go}(\gl):= \left( h^{(m)}(\gl) -\hat h _{N, \go} (\gl) \right)/\hat h _{N, \go} (\gl)$,
for the value $m=\hat m_{N,\go}$ explained in the caption of Fig.~\ref{fig:sec5_1} and
for the cases of $N=2.5\cdot 10^5$ ($\times$ dots), and $N=3.2\cdot 10^7$ ($+$ dots).
Notice that in the binary case the error is more important for small values of $\gl$ (recall
Table~\ref{tbl:2} and Fig.~\ref{fig:NCA}). Instead for the Gaussian case there is a deviation
both for small and large values of $\gl$: the deviation for large values is due to
the saturation effect explained in the text. Given the fact that $h_{\text{sat}}$, cf. \eqref{eq:sat},
behaves almost surely and to leading order for $N \to \infty$ as $\sqrt{\log N}$ one
understand  why the slow disappearing of the saturation effect has to be expected.
In both graphs the dotted line above the axis is at level $0.01$.
The fitted values for $\hat m_{N, \go}$, $N=  2.5\cdot 10^5$, are $0.821$ in the binary case
and $0.778$ in the Gaussian case.
}
\end{figure}

What we have observed numerically, see Figures~\ref{fig:sec5_1} and \ref{fig:sec5_2},
may be summed up by
the statement
\begin{equation}
\label{eq:cnj}
\text{there exists } m \text{ such that }
\hat h _{N, \go} (\gl) \, \approx h^{(m)}(\gl).
\end{equation}
Practically this means that $\hat h _{N, \go} (\gl)$, for a set of $\gl$
ranging from $0.05$ to $4$, may be fitted with remarkable precision
by the one parameter family of functions $\left\{ h^{(m)}(\cdot)\right\}_m$.
The fitting value of $m=: \hat m_{N, \go}$ does depend on $N$ and it is essentially increasing.
This is of course expected since localization requires a sufficiently large
system (recall in particular Table~\ref{tbl:2} and Fig.~\ref{fig:NCA} -- see the caption of
Fig.~\ref{fig:sec5_1} for the fitting criterion).
We stress that we are presenting results that have been obtained for one
fixed sequence of $\go$: based on what we have observed for example
in Section~\ref{sec:superadd} for different values of $\gl$ one does expect that
for smaller values of $\gl$ one should use larger values of $N$, but
 changing $N$ corresponds to selecting
a longer, or shorter, stretch of $\go$, that is a different sequence of charges
and this may have a rather strong effect on the value of $\hat m_{N, \go}$.
Moreover there is the problem of deciding which $\gl$-dependence to choose. This may explain
the deviations from \eqref{eq:cnj} that are observed for small values of $\gl$, but these
are in any case rather moderate (see Fig.~\ref{fig:sec5_2}).
\smallskip

A source of stronger (and unavoidable) deviations arises in the cases of unbounded charges:
of course if
\begin{equation}
\label{eq:sat}
h\, \ge \,  h_{\text{sat}} \, :=\, \max_{n\in \{1, \ldots, N \}} \left( -(\go_{2n-1}+ \go_{2n})/2 \right),
\end{equation}
then
$Z^{\gl, h}_{2N, \go} (0) <1$, regardless of the value of $\gl$.
Moreover it is immediate to verify that $\lim_{\gl \to \infty } Z^{\gl, h}_{2N, \go}(0) =+\infty$
for $h<  h_{\text{sat}}$ and therefore $\hat h _{N, \go} (\gl)\nearrow h_{\text{sat}} $
as $\gl \nearrow \infty$. We refer to the captions of Fig.~\ref{fig:sec5_2} for
more on this saturation effect.
 \smallskip

We have tried also alternative definitions of $\hat h _{N, \go} (\gl)$, namely:
\begin{enumerate}
\item the value of $h$ such that $ Z_{2N, \go}^{\gl, h} =1$ (or a different fixed value);
\item the value of $h$ such that the $\ell_1$ distance between
the distribution of the endpoint and the distribution of the meander, cf. Section~\ref{sec:path}, is smaller than a fixed threshold, for example $0.05$.
\end{enumerate}

What we have observed is that \eqref{eq:cnj} still holds.
What is not independent of the criterion is
 $\hat m_{N, \go}$.
 Of course believing deeply in \eqref{eq:cnj} entails
 the expectation  that $\hat m_{N, \go}$ converges to the non random
 quantity $h ^\prime_c (0)$.
  The results reported in this section suggest a value of $h ^\prime_c (0) $ larger than $0.83$
 and the cases presented in Section~\ref{sec:path} suggest that it should be smaller than $0.86$.


\appendix

\section{The algorithm for computing $Z_{N,\go}$}
\label{app:algo}

We are going to briefly illustrate the algorithm we used in the numerical computation of the partition function $Z_{N}=Z_{N,\go}^{\gl,h}$. We recall its definition (see equation (\ref{eq:Boltzmann})):
\begin{equation} \label{eq:appZ}
Z_N = \bE \Bigg[ \exp \Bigg( -2\gl \sum_{n=1}^N (\go_n + h) \gD_n \Bigg) \Bigg]\,,
\end{equation}
where $\gD_n := (1-\sign(S_n))/2$ and the convention for $\sign(0)$ described in the introduction.

Observe that a direct computation of $Z_N$ from (\ref{eq:appZ}) would require to sum the contributions of $2^N$ random walk trajectories, making the problem numerically intractable.
 However, here we can make profitably use of the \textsl{additivity} of our Hamiltonian:
 loosely speaking, if we join together two (finite) random walk segments, the energy of the resulting path is the sum of the energies of the building segments.

We can exploit this fact to derive a simple recurrence relation for the sequence of functions $\big\{ \cZ _{M}(y):=Z_{2M}(2y),\ y \in \Z\big\}_{M\in\N}$, where $Z_N(x) = Z_{N,\go}^{\gl,h}(x)$, the latter
defined in \eqref{eq:pinned},
and we recall that we work  with even values of~$N$. Conditioning on $S_{2M}$ and using the Markov property one easily finds
\begin{equation} \label{eq:apprec}
\cZ_{M+1} (y) = \begin{cases}
\frac14 \cZ_{M}(y+1) \;+\; \frac12 \cZ_{M}(y) \;+\; \frac14 \cZ_{M}(y-1) & y>0 \\
\frac14 \Big[ \cZ_{M}(1) + \cZ_{M}(0) \Big] \;+\; \frac14 \ga_M \Big[ \cZ_{M}(0) + \cZ_{M}(-1) \Big] & y=0 \\
\ga_M \Big[ \frac14 \cZ_{M}(y+1) \;+\; \frac12 \cZ_{M}(y) \;+\; \frac14 \cZ_{M}(y-1) \Big] & y<0
\end{cases} \;,
\end{equation}
where we have put $\ga_M := \exp\big(-2\gl\,(\go_{2M+1} + \go_{2M+2} + 2h)\big)$.

From equation (\ref{eq:apprec}) and from the trivial observation that $\cZ_{M}(y)=0$ for $|y|>M$, it follows that $\{\cZ_{M+1}(y),\ y \in \Z\}$ can be obtained from $\{\cZ_{M}(y),\ y \in \Z\}$ with $O(M)$ computations. This means that we can compute $Z_N$  in $O(N^2)$ steps.\footnote{The algorithm just described can be implemented in a standard way: the code we used, written in  C,
 is available on the web page:
 {\tt http://www.proba.jussieu.fr/pageperso/giacomin/C/prog.html}. Graphic representations and standard statistical procedures have been performed
with R \cite{cf:R}. }

\medskip

We point out that sometimes one is satisfied with \textsl{lower bounds} on $Z_N$, for instance in the statistical text for localization described in Section~\ref{sec:superadd}. In this case the algorithm can be further speeded up by restricting the computation to a suitable set of random walk trajectories. In fact when the system size is~$N$ the polymer is at most at distance $O(\sqrt{N})$ (we recall the discussion in Section~\ref{sec:path} on the path behavior), hence a natural choice to get a lower bound on $Z_N$ is to only take into account the contribution coming from those random walk paths $\{s_n\}_{n\in\N}$ for which
\begin{equation}
-A\sqrt{n} \le s_n \le B \sqrt{n} \qquad \text{for } n \ge N_0\,,
\end{equation}
where $A,B,N_0$ are positive constants. Observe that this is easily implemented in the algorithm described above: it suffices to apply relation (\ref{eq:apprec}) only for $y\in[-A\sqrt M, B\sqrt M]$, while setting $\cZ_{M+1}(y)=0$ for the other values of~$y$. In this way the number of computations needed to obtain $Z_N$ is reduced to~$O(N^{3/2})$.

The specific values of $A,B,N_0$ we used in our numerical computations are $3,8,1000$, and we would like to stress that the lower bound on $Z_N$ we got coincides up to the $8^{\text{th}}$ decimal digit with the {\sl true value} obtained applying the complete algorithm.


\smallskip
A final important remark is that for the results we have reported we have used
the Mersenne--Twister~\cite{cf:MT} pseudo--random number generator.
However we have also tried other pseudo--random number generators
and {\sl true randomness} from {\tt www.random.org}:
the results appear not to depend on the generator.


\section{Proof of the lower bound on $h_c$}
\label{app:prooflb}
We are going to give a detailed proof of the lower bound \eqref{eq:mainBG} on the critical curve, together with some related result. We stress that this appendix can be made substantially lighter
if one is interested only in the {\sl if} part of Proposition~\ref{prop:stopping1}. In this case
the first part of this appendix is already contained in the first part of \S~\ref{sec:lb_outline}, up
to \eqref{eq:forappB}, and it suffices to look at \S~\ref{sec:appB2}.

We recall that $Z_{N,\go}^{\gl,h}(0)$ is the partition function corresponding to the polymer pinned at its right endpoint, see \eqref{eq:pinned}, and $T^C=T^{C}(\go)$ is the first $N$ for which $Z_{N,\go}(0)\ge C$, see \eqref{eq:T^C}. In particular, for all $\go$ such that $T^C(\go) < \infty$ we have
\begin{equation} \label{eq:stopping_maj}
Z_{T^C(\go),\go}^{\gl,h}(0) \ge C\,.
\end{equation}
We will also denote by $\cF_n := \gs(\go_1,\ldots,\go_n)$ the natural filtration of the sequence $\{\go_n\}_{n\in\N}$.

\smallskip

\subsection{A different look at (de)localization}
We want to show that (de)localization can be read from $T^C$. We introduce some notation: given an increasing, $2\N$--valued sequence $\{t_i\}_{i\in\N}$, we set $t_0:=0$ and $\zeta_N:=\max\{k: t_k \le N\}$. Then we  define
\begin{align} \label{eq:low_b}
\begin{split}
\widehat{Z}_{N,\go}(0) = \widehat{Z}_{N,\go}^{\{t_i\},\gl,h}(0) & \;:=\; \bE \left[ e^{-2 \gl \sum_{n=1}^N \left( \go_n +h\right) \Delta_n}; \, S_{t_1}=0,\, \ldots,\, S_{t_{\zeta_N}}=0,\, S_N= 0 \right] \\
&\;=\; \prod_{i=0}^{\zeta_N-1} Z_{t_{i+1}-t_i,\theta^{t_i}\go}^{\gl,h}(0) \,\cdot\, Z_{N-t_{\zeta_N}(\go),\theta^{t_{\zeta_N}}\go}^{\gl,h}(0)\,,
\end{split}
\end{align}
and we recall that $\theta$ denotes the translation on the environment.
One sees immediately that $\widehat{Z}_{N,\go}(0)\le Z_{N,\go}(0)$.
We first establish a preliminary result.

\begin{lemma} \label{lem:chop}
If the sequence $\{t_i\}_i$ is such that $\zeta_N/N \to 0$ as $N\to\infty$, then
\begin{equation}
\lim_{N\to\infty} \frac 1N \log \widehat{Z}_{N,\go}^{\{t_i\},\gl,h}(0) \;=\; \tf(\gl,h)\,,
\end{equation}
both $\bbP(\dd\go)$--a.s. and in $\bbL_1(\bbP)$.
\end{lemma}

\proof
By definition we have $Z_{N,\go}(0) \ge \widehat{Z}_{N,\go}(0)$. On the other hand, we are going to show that
\begin{equation} \label{eq:rough_maj}
Z_{N,\go}^{\gl,h}(0) \;\le\; 4^{\zeta_N} \, A^{2\zeta_N} \, \left( \prod_{i=1}^{\zeta_N} (t_i-t_{i-1}) \cdot (N-t_{\zeta_N}) \right)^{3} \,  \widehat{Z}_{N,\go}^{\{t_i\},\gl,h}(0)\,,
\end{equation}
where $A$ is a positive constant. To derive this bound, we resort to the equation \eqref{eq:reducttoexc} that expresses $Z_{N,\go}(0)$ in terms of random walk excursions. We recall that $K(2n)$ is the discrete probability density of the first return time of the walk $S$ to $0$, and that $K(t) \ge 1/(A\,t^{3/2})$, $t\in 2\N$, for some positive constant~$A$: it follows that for $a_1, \ldots, a_k \in 2\N$
\begin{equation} \label{eq:app_bound}
K(a_1 + \ldots + a_k) \; \le \; 1 \; \le \; A^k \, (a_1 \cdot \ldots \cdot a_k)^{3/2} \, K(a_1) \cdot \ldots \cdot K(a_k)\,.
\end{equation}
This gives us an upper bound to the entropic cost needed to split a random walk excursion of length $(a_1 + \ldots + a_k)$ into $k$~excursions of lengths $a_1, \ldots, a_k$.

Now let us come back to the second line of \eqref{eq:reducttoexc}, that can be rewritten as
\begin{equation}
\label{eq:app_sum}
Z_{N,\go}(0) \,=\, \sum_{\{x_i\} \subseteq \{0, \ldots, N\} \cap 2\N} G(\{x_i\})\,.
\end{equation}
A first observation is that if we restrict the above sum  to the $\{x_i\}$ such that
$\{x_i\} \supseteq \{t_i\}$, then we  get  $\widehat{Z}_{N,\go}^{\{t_i\}}(0)$. Now for each $\{x_i\}$ we
aim at finding an  upper bound on the term $G(\{x_i\})$ of the form
 $c \cdot G(\{x_i\} \cup \{t_i\})$ for some $c>0$ not depending on $\{x_i\}$.
 Each term $G(\{x_i\})$, see \eqref{eq:reducttoexc}, is the product of two terms: an entropic part depending on~$K(\cdot)$ and an energetic part depending on~$\varphi(\cdot)$. Replacing the entropic part costs no more than
\begin{equation}
    c_{\text{ent}} \, :=\,  A^{2\zeta_N} \, \left( \prod_{i=1}^{\zeta_N} (t_i-t_{i-1}) \cdot (N-t_{\zeta_N}) \right)^{3}\,,
\end{equation}
thanks to~\eqref{eq:app_bound}. On the other hand, the cost for replacing the energetic part is easily
bounded above by
\begin{equation}
    c_{\text{energy}} \, :=\, 2^{\zeta_N}\,,
\end{equation}
so that the bound $G(\{x_i\}) \le c \cdot G(\{x_i\} \cup \{t_i\})$ holds true with $c:= c_{\text{ent}} \, c_{\text{energy}}$. Replacing in this way each term in the sum in the r.h.s. of \eqref{eq:app_sum}, we are left with a sum of terms $G(\{y_i\})$ corresponding to sets $\{y_i\}$ such that $\{y_i\} \supseteq \{t_i\}$. It remains to count the {\sl multiplicity} of any such~$\{y_i\}$, that is how many original sets $\{x_i\}$ are such that $\{x_i\} \cup \{t_i\} = \{y_i\}$. Sets $\{x_{i}\}$ satisfying this last condition must differ only for a subset of $\{t_{i}\}$, hence the sought multiplicity is $2^{\zeta_N}$ (the cardinality of the parts of $\{t_{i}\}$) and the bound \eqref{eq:rough_maj} follows.

Therefore we get
\begin{align}
\bigg| \frac{\log \widehat{Z}_{N,\go}^{\{t_i\},\gl,h}(0)}{N}  -  \frac{\log Z_{N,\go}^{\gl,h}(0)}{N} \bigg| & \;\le\; (2\log 2 A) \frac{\zeta_N}{N} \;+\; 3\, \frac{1}{N} \,\log \left( \prod_{i=1}^{\zeta_N} (t_i-t_{i-1}) \cdot (N-t_{\zeta_N}) \right) \\
&\;\le\; (2\log 2A) \frac{\zeta_N}{N} \;+\; 3 \, \frac{\zeta_N+1}{N} \, \log \bigg(\frac{N}{\zeta_N+1}\bigg)\,,
\nonumber
\end{align}
where in the second inequality we have made use of the elementary fact that once the sum of $k$ positive numbers is fixed, their product is maximal when all the numbers coincide (for us $k=\zeta_N+1$). Since by hypothesis $\zeta_N/N \to 0$ as $N\to\infty$, the Lemma is proved.\qed

\bigskip

Now we are ready to prove the characterization of $\cL$ and $\cD$ in terms of $T^C$.
Fix any $C>1$.

\smallskip

\begin{proposition} \label{prop:stopping1}
A point $(\gl,h)$ is localized, that is $h < h_c(\gl)$, if and only if $\bbE[T^C]<\infty$.
\end{proposition}

\proof
We set $\cA:=\{\go: T^C(\go)<\infty\}$. Observe that for $\go\in \cA^{\complement}$ we have $Z_{N,\go}(0) \le C$ for every $N\in2\N$, and consequently $\log Z_{N,\go}^{\gl,h}(0)/N \to 0$ as $N\to\infty$.

Consider first the case when the random variable $T^C$ is defective, that is $\bbP[\cA^{\complement}]>0$ (this is a particular case of $\bbE[T^C]=\infty$). Since we know that $\log Z_{N,\go}^{\gl,h}(0)/N \to \tf(\gl,h)$, $\bbP(\dd\go)$--a.s., from the preceding observation it follows that $\tf(\gl,h)=0$ and the Proposition is proved in this case.

\smallskip

Therefore in the following we can assume that $T^C$ is proper, that is $\bbP(\cA)=1$, so that equation (\ref{eq:stopping_maj}) holds for almost every~$\go$. Setting $\theta^{-1}\cA := \{\go : \theta\go \in \cA\}$, we have $\bbP\left(\theta^{-1}\cA\right)=1$ since $\bbP$ is $\theta$--invariant, and consequently $\bbP\left(\cap_{k=0}^\infty \theta^{-k}\cA
 \right)=1$, which amounts to saying that (\ref{eq:stopping_maj}) can be actually strengthened to
\begin{equation} \label{eq:im_stopping_maj}
Z_{T^C(\theta^k\go),\theta^k\go}^{\gl,h}(0) \ge C \qquad \forall k \ge 0,\ \bbP(\dd\go)\text{--a.s.}\,.
\end{equation}
Observe that the sequence $\{( \theta^{T^C(\go)} \go )_n\}_{n\in\N}$ has the same law as $\{\go_n\}_{n\in\N}$ and it is independent of $\cF_{T^C}$. We can define inductively an increasing sequence of stopping times $\{T_n\}_{n\in\N}$ by setting $T_0:=0$ and $T_{k+1}(\go) - T_k(\go) := T^C(\theta^{T_k(\go)}\go) =: S_k(\go)$. We also set $\zeta_N(\go) := \max \{n:\ T_n(\go) \le N\}$. Since $\{S_k\}_{k\in\N}$ is an IID sequence, by the strong law of large numbers we have that, $\bbP(\dd\go)$--a.s., $T_n(\go)/n \to \bbE[T^C]$ as $n\to\infty$, and consequently $\zeta_N(\go)/N \to 1/\bbE[T^C]$ as $N\to\infty$ (with the convention that $1/\infty=0$).

Now let us consider the lower bound $\widehat{Z}_{N,\go}(0)$ corresponding to the sequence $\{t_i\}=\{T_i(\go)\}$: from \eqref{eq:low_b} and \eqref{eq:im_stopping_maj} we get that $\bbP(\dd\go)$--a.s.
\begin{align} \label{eq:major}
\begin{split}
\widehat{Z}_{N,\go}^{\{T_i(\go)\},\gl,h}(0) &\;=\; \prod_{i=0}^{\zeta_N(\go)-1} Z_{T^C(\theta^{T_i}\go),\theta^{T_i}\go}^{\gl,h}(0) \,\cdot\, Z_{N-T_{\zeta_N(\go)}(\go),\theta^{T_{\zeta_N(\go)}}\go}^{\gl,h}(0) \\
&\;\ge\; C^{\zeta_N(\go)} \cdot \frac{c}{N^{3/2}}\,,
\end{split}
\end{align}
where $c$ is a positive constant (to estimate the last term we have used the  lower bound $Z_k(0) \ge c/k^{3/2}$, cf. \eqref{eq:step_deloc}), and consequently
\begin{equation}
\tf(\gl,h) \;=\; \lim_{N\to\infty} \frac{\log Z_{N,\go}^{\gl,h}(0)}{N} \;\ge\; \liminf_{N\to\infty} \frac{\log \widehat{Z}_{N,\go}^{\{T_i(\go)\},\gl,h}(0)}{N} \;\ge\; \frac{\log C}{\bbE[T^C]}\,.
\end{equation}
It follows that if $\bbE[T^C]<\infty$  then $\tf(\gl,h)>0$, that is $(\gl,h)$ is localized.

\smallskip

It remains to consider the case $\bbE[T^C]=\infty$, and we want to show that this time $\widehat{Z}_{N,\go}(0)$, defined in \eqref{eq:major}, gives a null free energy. In fact, as $T^C(\eta)$ is defined as the \textsl{first} $N$ such that $Z_{N,\eta}(0) \ge C$, it follows that $Z_{T^C(\eta),\eta}(0)$ cannot be much greater than $C$. More precisely, one has that
\begin{equation}
Z_{T^C(\eta),\eta}(0) \le  C \,\exp(2\gl|\eta_{T^C(\eta)-1} + \eta_{T^C(\eta)}|)\,,
\end{equation}
and from the first line of \eqref{eq:major} it follows that
\begin{equation}
\frac 1N \log \widehat{Z}_{N,\go}(0) \;\le\; \frac{\zeta_N(\go)+1}{N} \log C \;+\;  \frac{2\gl}{N} \sum_{i=1}^{\zeta_N(\go)} \Big( |\go_{T_i(\go)}| + |\go_{T_i(\go)-1}| \Big)\,.
\end{equation}
We estimate the second term in the r.h.s. in the following way:
\begin{align}
    & \frac 1N \sum_{i=1}^{\zeta_N(\go)} \Big( |\go_{T_i(\go)}| + |\go_{T_i(\go)-1}| \Big) = \frac 1N \sum_{k=1}^{N} \ind_{\{\exists i:\, T_i(\go) = k\}} \Big( |\go_k| + |\go_{k-1}| \Big)\nonumber  \\
    & \qquad \le \left( \frac 1N \sum_{k=1}^{N} \ind_{\{\exists i:\, T_i(\go) = k\}} \right) ^{1/2} \left( \frac 1N \sum_{k=1}^{N} \Big( |\go_k| + |\go_{k-1}| \Big)^2 \right)^{1/2}\\
    &\qquad \le \sqrt{\frac{\zeta_N(\go)}{N}} \cdot 2 \sqrt{ \frac 1N \sum_{k=1}^{N} |\go_k|^2} \le A \sqrt{\frac{\zeta_N(\go)}{N}} \,, \nonumber
\end{align}
for some positive constant~$A=A(\go)$ and  eventually as $N \to \infty$, having used the Cauchy--Schwarz inequality and the law of large numbers for the sequence $\{|\go_k|^2\}_{k\in\N}$. Therefore
\begin{equation}
    \frac 1N \log \widehat{Z}_{N,\go}(0) \;\le\; \frac{\zeta_N(\go)+1}{N} \log C \;+\; 4\gl A \sqrt{\frac{\zeta_N(\go)}{N}}\,,
\end{equation}
and since $\bbE[T^C]=\infty$ implies $\zeta_N(\go)/N \to 0$, $\bbP(\dd\go)$--a.s., we have $\log \widehat{Z}_{N,\go}(0) /N \to 0$, $\bbP(\dd\go)$--a.s.. Then Lemma~\ref{lem:chop} allows us to conclude that $\tf (\gl,h)=0$, and the proof of the Proposition is completed.\qed


\smallskip

\subsection{Proof of the lower bound on $h_c$}
\label{sec:appB2}
To prove equation \eqref{eq:mainBG}, we are going to build, for every $(\gl,h)$ such that $h < \underline h(\gl)$, a random time $T$ such that $\bbE[T]<\infty$ and $Z_{T(\go),\go}^{\gl,h}(0) \ge C$, for some $C>1$. It follows that $T^C \le T$, yielding that $\bbE[T^C]<\infty$ and by Proposition~\ref{prop:stopping1} $(\gl,h)$ is localized, that is, $\underline h(\gl) \le h_c(\gl)$.

\smallskip

Given $M\in 2\N$
and $q<-h$, we start defining the stopping time
\begin{equation}
\tau_M(\go) = \tau_{M,q}(\go) := \inf \bigg\{n\in2\N:\ \exists k \in 2\N,\ k \ge M:\ \frac{\sum_{i=n-k+1}^n \go_i}{k} \le q \bigg\}\,.
\end{equation}
This is the first instant at which a $q$--atypical stretch of length at least $M$ appears along the sequence $\go$. The asymptotic behavior of $\tau_M$ is given by Theorem 3.2.1 in \cite[\S~3.2]{cf:DZ} which says that $\bbP(\dd\go)$--a.s.
\begin{equation} \label{eq:as_tau}
\frac{\log \tau_{M}(\go)}{M} \to \gS(q) \qquad \text{as } M\to\infty\,,
\end{equation}
where $\gS(q)$ is Cramer's Large Deviations functional for $\go$, \eqref{eq:cramer}.
We also give a name to the shortest of the terminal stretches in the definition of $\tau_M$:
\begin{equation}
R_{M}(\go) = R_{M,q}(\go) := \inf \bigg\{k \in 2\N,\ k \ge M:\ \frac{\sum_{i=\tau_M-k+1}^{\tau_M} \go_i}{k} \le q \bigg\}\,,
\end{equation}
and it is not difficult to realize that $R_M \le 2M$.

\smallskip

We are ready to give a simple lower bound on the partition function of size $\tau_{M,q}$ (for any $M\in 2\N$ and $q<-h$): it suffices to consider the contribution of the trajectories that are negative in correspondence of the last (favorable) stretch of size $R_M$, and stay positive the rest of the time. Recalling that
we use $K(\cdot)$ for the discrete density of the first return time to the origin and that by
\eqref{eq:asympt} we have $K(2n)\ge c/n^{-3/2}$ for a constant $c>0$,
we estimate
\begin{equation}\begin{split}
Z_{\tau_M(\go), \go}^{\gl,h}(0) & \ge \frac 1 4 \, K\left({\tau_M - R_M}\right) \, K\left(
{R_M}\right) \, e^{-2\gl (q+h) R_M} \ge \frac{c^2}{4 \tau_M^{3/2} (2M)^{3/2}} e^{-2\gl (q+h) M} \\
& \ge c'\, \exp \bigg\{ \frac 3 2 M \bigg[ (-4\gl /3)q - \frac{\log \tau_M}{M} - ( 4 \gl/3) h - \frac{\log M}{M} \bigg] \bigg\} \,,
\end{split}\end{equation}
where $c' := c^2/(8\sqrt{2})$.

Having in mind (\ref{eq:as_tau}), we define a random index $\ell=\ell_{A,\gep,q}$ depending on the two parameters $A \in 2\N,\ \gep>0$ and on $q$:
\begin{equation} \label{eq:def_ell}
\ell(\go)  = \ell_{A,\gep,q}(\go) := \inf \bigg\{ k\in 2 \N,\ k \ge A:\ \frac{\log \tau_{k,q}(\go)}{k} \le \gS(q) + \gep \bigg\} \,,
\end{equation}
and we finally set $T (\go) = T_{A,\gep,q} (\go) := \tau_{\ell(\go)}(\go)$. Then for the partition function of size $T(\go)$ we get
\begin{equation} \label{eq:almost_maj}
Z_{T(\go), \go}^{\gl,h}(0) \ge c'\, \exp \bigg\{ \frac 3 2 A \bigg[ (-4\gl /3)q - \gS(q) - (4\gl /3)h - \frac{\log A}{A} - \gep \bigg] \bigg\} \,.
\end{equation}

The fact that $\bbE[T_{A,\gep,q}]<\infty$ for any choice of $A,\gep,q$ (with $q<-h$) is proved in Lemma~\ref{lem:stopping2} below. It only remains to show that for every fixed $(\gl,h)$ such that $h<\underline h(\gl)$, or equivalently
\begin{equation}
\label{eq:cond_low}
(4\gl/3)h < \log \M(-4\gl/3)\,,
\end{equation}
the parameters $A,\gep,q$ can be chosen such that the right--hand side
 of equation (\ref{eq:almost_maj}) is greater than~$1$.

\smallskip

The key point is the choice of  $q$. Note that
the generating function $ \M(\cdot)$ is smooth, since  finite on the whole real line.
 Moreover for all $ \gl \in\R$ there exists some $q_0\in\R$ such that
\begin{equation}
\log \M(-4\gl/3) = (-4\gl/3) q_0 - \gS(q_0)\,,
\end{equation}
and from (\ref{eq:cond_low}) it follows that $q_0 < -h$. Therefore we can take $q=q_0$, and equation (\ref{eq:almost_maj}) becomes
\begin{equation}
\label{eq:lbonZTom}
Z_{T(\go), \go}^{\gl,h}(0) \ge c'\, \exp \bigg\{ \frac 3 2 A \bigg[ \log \M(-4\gl/3) - (4\gl/3) h - \frac{\log A}{A} - \gep \bigg] \bigg\} \,.
\end{equation}
It is now clear that for every $(\gl,h)$, such that (\ref{eq:cond_low}) holds, by choosing
$\gep$ sufficiently small and $A$ sufficiently large, the right--hand side of \eqref{eq:lbonZTom} is greater than~1, and the proof of \eqref{eq:mainBG} is complete.

\medskip

\begin{lemma}
\label{lem:stopping2}
For every $A\in 2\N$,  $\gep > 0$ and $ q <-h$
the random variable $T(\go)=T_{A,\gep,q}(\go)$ defined
below \eqref{eq:def_ell} is integrable: $\bbE[T]<\infty$.
\end{lemma}

\medskip

\proof

By the definition (\ref{eq:def_ell}) of $\ell=\ell_{A,\gep,q}$ we have
\begin{equation}
T_{A,\gep,q} \le \exp\big( (\gS(q) + \gep)\, \ell_{A,\gep,q} \big)\,,
\end{equation}
so it suffices to show that for any $\gb>0$ the random variable $\exp(\gb\, \ell_{A,\gep,q})$ is integrable.

For any $l\in2\N$, we introduce the IID sequence of random variables $\{Y_n^{l}\}_{n\in\N}$ defined by
\begin{equation}
Y^{l}_n := \frac 1 {l}{\sum_{i=(n-1)l+1}^{n l} \go_i}\,.
\end{equation}
By Cramer's Theorem \cite{cf:DZ} we have that for any fixed $q<0$ and $ \gep>0$ there exists  $l_0$ such that $\bbP\left(Y_1^{l} \le q\right) \ge e^{-l (\gS(q) + \gep/2)}$ for every $l\ge l_0$.
By (\ref{eq:def_ell}) have that
\begin{equation}
\begin{split}
\{\ell > l\} \subseteq \left\{\tau_l > \exp((\gS(q)+\gep) l)\right\}
\subseteq \inter_{i=1}^{\lfloor M/l \rfloor} \{Y_i^{l} > q\}\,,
\end{split}\end{equation}
with $M:=\exp((\gS(q)+\gep) l)$, so that
\begin{equation}
\begin{split}
\bbP\left(\ell > l\right)
  \le \left( 1- e^{-l (\gS(q) + \gep/2)} \right)^{\lfloor M/l \rfloor}
  &\le
  \exp \left( - \lfloor M/l \rfloor e^{-l (\gS(q) + \gep/2)} \right)
  \\
& \le  \exp \left(- \exp\left(l \gep/4\right)\right),
\end{split}
\end{equation}
where the last step holds if $l$ is sufficiently large   (we have also used
 $1-x \le e^{-x}$). Therefore
\begin{equation}
\bbP\left(\exp(\gb\,\ell) > N\right) = \bbP\left(\ell > (\log N) /\gb \right)
\le  \exp \left( - N ^{\gep / 4\gb }\right),
\end{equation}
when $N$ is large, and the proof is complete.

\qed

\bigskip


\section*{Acknowledgements}

We thank Thierry Bodineau, Erwin Bolthausen and Fabio Toninelli 
 for very useful discussions.
We are also very grateful to Jacques Portes for the constant hardware assistance
during the development of this work.


\end{document}